\documentclass[a4paper,fleqn,usenatbib]{mnras}

\usepackage[T1]{fontenc}
\usepackage{ae,aecompl}

\usepackage{amssymb}	
\usepackage{amsmath}
\usepackage{graphicx}
\usepackage[usenames]{color}
\usepackage{bm}

\newcommand{\Ddel}{\delta_{\rm D}   }

\newcommand{\MpcOh}{ \,  \mathrm{Mpc}  \, h^{-1} }
\newcommand{\hOMpc}{ \,  \mathrm{Mpc}^{-1}  \, h  }
\newcommand{\nn}{ \nonumber }
\newcommand{\Msun}{ \,   M_{\odot}  {h}^{-1}   }

\newcommand{\beq}{\begin{equation}}
\newcommand{\eeq}{\end{equation}}

\newcommand{\comment}[1]{}

\title[Consistency relations and implications]{Consistency relations for the Lagrangian halo bias and their implications}

\author[K. C. Chan, R. K. Sheth, and R. Scoccimarro]{
Kwan Chuen Chan,$^{1}$\thanks{E-mail: chan@ice.cat (KCC)}
Ravi K. Sheth,$^{2}$
Rom\'an Scoccimarro$^{3}$
\\
$^{1}$ Institute of Space Sciences, IEEC-CSIC, Campus UAB, Carrer de Can Magrans, s/n,  08193 Bellaterra, Barcelona, Spain       \\
$^{2}$ Center for Particle Cosmology, University of Pennsylvania, 209 S. 33rd St., PA 19104,  Philadelphia, USA   \\
$^{3}$ Center for Cosmology and Particle Physics, Department of Physics, 
 New York University, NY 10003, New York, USA
}

\date{Accepted XXX. Received YYY; in original form ZZZ}

\pubyear{2016}

\begin{document}
\label{firstpage}
\pagerange{\pageref{firstpage}--\pageref{lastpage}}
\maketitle

\begin{abstract}
  The protohalo patches from which halos form are defined by a number of constraints imposed on the Lagrangian dark matter density field.  Each of these constraints contributes to biasing the spatial distribution of the protohalos relative to the matter.  We show how measurements of this spatial distribution -- linear combinations of protohalo bias factors -- can be used to make inferences about the physics of halo formation.  Our analysis exploits the fact that halo bias factors satisfy consistency relations which encode this physics, and that these relations are the same even for sub-populations in which assembly bias has played a role.  We illustrate our methods using a model in which three parameters matter:  a density threshold, the local slope and the curvature of the smoothed density field.  The latter two are nearly degenerate; our approach naturally allows one to build an accurate effective two-parameter model for which the consistency relations still apply.  This, with an accurate description of the smoothing window, allows one to describe the protohalo-matter cross-correlation very well, both in Fourier and configuration space.  We then use our determination of the large scale bias parameters together with the consistency relations, to estimate the enclosed density and mean slope on the Lagrangian radius scale of the protohalos.  Direct measurements of these quantities, made on smaller scales than those on which the bias parameters are typically measured, are in good agreement.
\end{abstract}




\section{Introduction}

Halos and the galaxies they host are biased tracers of the dark matter density field \citep{Kaiser84, BBKS86,MoWhite1996}. To extract cosmological information from galaxy surveys, this bias must be understood.  
In the best studied models of halo formation, the bias is a consequence of the fact that the protohalo patches from which halos form are defined by a number of constraints imposed on the initial Lagrangian dark matter density field.  Each one of these constraints contributes to biasing the spatial distribution of the protohalos relative to the matter.  Thus, in principle, there is valuable halo formation physics hidden in the bias parameters. In fact, the bias parameters satisfy a hierarchy of consistency relations \citep{MussoParanjapeSheth2012,ParanjapeShethDesjacques2013}. These consistency relations not only allow us to check the self-consistency of the bias prescription, but they also potentially open the road to learning about the physics of halo formation. The main goal of this paper is to demonstrate that we can indeed extract information about the small scale physics of halo formation from measurements of the large scale clustering of halos.  In particular, our methodology allows one to estimate if assembly bias effects, of the sort first identified by \cite{ShethTormen2004}, are present in the halo population.  In this respect, our work complements that of \cite{CastorinaParanjapeSheth2016, CastorinaParanjapeHahnSheth2016}; whereas they used configuration space methods to address similar issues, we, like \cite{Modi:2016dah}, use Fourier space measurements.  

This paper is organized as follows. We provide a new derivation of the consistency relations for the linear Lagrangian bias parameters using a straightforward linear algebra method in Sec.~\ref{sec:general}.  Since protohalos are extended objects, we discuss the importance of smoothing in Sec.~\ref{sec:Ws}.  In Sec.~\ref{sec:specific} we illustrate our arguments with some specific examples, which we use to address the question of degeneracies between parameters and effective versus exact models of the physics.  After showing the measurements of the correlations in configuration and Fourier space in Sec.~\ref{sec:correlation_measurements}, in Sec.~\ref{sec:direct} we discuss our direct estimates of quantities which are thought to matter for the small scale physics of halo formation, some of which are novel. In Sec.~\ref{sec:crosspowerspectrum}, we use a two-bias parameter model to fit measurements of the Fourier space bias signal. The halo collapse threshold inferred from the consistency relation is compared with the direct measurements of the overdensity within Lagrangian protohalos in Sec.~\ref{sec:consist_rel_density}, and a similar test of the profile slope is in Sec.~\ref{sec:consistency_rel_firstcorssing}.  We revisit the physical meaning of the consistency relations in Sec.~\ref{sec:revisit}.  We summarize our findings and conclude in Sec.~\ref{sec:conclusion}. An Appendix is devoted to the study of the correlation function in real space, and shows that although bias parameters may depend on smoothing window, the combination which matters for the consistency relation does not.  

\section{Lagrangian constraints, bias and consistency relations}
\label{sec:Theory}

Although ultimately  we are interested in halos in Eulerian space, as they are potentially observable, the modelling of halo properties often starts in Lagrangian space.  This is primarily because the statistics of the Lagrangian field are easier to describe, particularly if the initial conditions were Gaussian.  The best-studied models are the peak \citep{Kaiser84, BBKS86, DesjacquesSheth2010}, excursion set \citep{PressSchechter,BCEK1991,MussoParanjapeSheth2012} and excursion set peak \citep{AppelJones1990,ParanjapeSheth2012, ParanjapeShethDesjacques2013,BiagettiChanetal_2014,Dizgah:2015kqi} approaches.  In all three approaches, describing how the initial Lagrangian protohalos evolve to form the final Eulerian halos is a separate step.  In what follows, we do not consider this second step, except to point out that the way forward is described in \cite{DesjacquesCrocceetal2010}.  For a recent review on halo bias, see \cite{Desjacques:2016bnm}. 


\subsection{ General formalism}\label{sec:general}
In one of the simplest models, a protohalo is identified with any position where the smoothed field exceeds a (physically motivated) threshold.  The peak model adds the additional constraints that the spatial gradient of the smoothed field should vanish and that the curvature should be negative.  These constraints on the scale of the protohalo patch impact correlations between the protohalo centers and the large-scale matter distribution.  Our goal is to extract these constraints from large-scale cross correlations.   

However, to do so, we must first address the fact that `smoothing' is common to all halo formation models.  I.e., it is the average properties of the field centered on a patch in the initial conditions which determine whether or not it will become a halo.  Therefore, the shape of the smoothing window $W$ is expected to play an important role.  For this reason, it is important to note that the analysis which follows is generic for all window choices; we will only specify our choice of window in the next subsection.  

We suppose that the constraints are given by the vector $C$.  If the constraint variables are set to be  $C = \mathcal{C}$, then the expectation value of the large-scale field $\Delta$ given the constraints is 
\beq
\langle \Delta |C =  \mathcal{C}  \rangle =  \frac{  \int d \Delta  \,   p( \Delta | C=\mathcal{C} )   \Delta     }{  \int d \Delta  \,  p( \Delta | C=\mathcal{C} )    }, 
\eeq
where $p(\Delta | C=\mathcal{C})$ is the distribution of $ \Delta $ conditioned on $C=\mathcal{C}$.  In general, we must integrate over the distribution of the constraint $\Pi(C)$ as well:
\beq
\label{eq:Delta_averageC}
\langle \Delta |C \rangle =  \frac{ \int d \mathcal{C}  \,  \Pi ( \mathcal{C} )    \int d \Delta  \,   p( \Delta |  \mathcal{C} )  \,  \Delta   }{  \int d \mathcal{C}   \,   \Pi ( \mathcal{C} )  \int  d \Delta  \,  p( \Delta |  \mathcal{C} )  }. 
\eeq

We will often assume that $ \Delta $ is the density field smoothed on a large scale at the same spatial position at which the constraints were specified, but we can take it to be other fields at other positions if we wish.  If  $ \Delta $ is taken to be the density, $ \langle \Delta | C \rangle $ would be the profile of the Lagrangian halos.  The large-scale field only serves as a surrogate for extracting the smaller scale halo formation physics.  This physics is encoded in the constraint $ C = (C_1, \dots, C_n)$, which we express in terms of normalized (zero-mean unit variance) random variables. For example, it can be $( \nu, x, \dots  )$, where $\nu$ is the peak height and $x$ its curvature.  The Fourier transforms of these quantities are defined as 
\begin{align}
\label{eq:nu_def}
 \nu( \bm{k}  ) &= \frac{ W(kR) }{ \sqrt{ s_0 } } \, \delta_{\rm m}(\bm{k})  , \\
\label{eq:x_def}
   x( \bm{k}) &= \frac{ k^2 W(kR)  }{  \sqrt{ s_2 } } \, \delta_{\rm m}(\bm{k}) , 
\end{align}
where  $\delta_{\rm m} (\bm{k}) $ is (a Fourier mode of) the dark matter density contrast, and $W$ is the smoothing window. The spectral moment $s_j$ is defined as
\beq
 \label{eq:sj}
 s_j(R) \equiv  \int \frac{ dk}{2 \pi^2}  \,  k^{ 2( 1+j) } P_{\rm m}(k)  W^2 ( kR ), 
\eeq
where $P_{\rm m} (k)$ is the linear power spectrum evaluated at $z=0$.  In what follows, we will also be interested in the `slope' variable
\beq
 \label{eq:u_def}
 u ( \bm{k} ) = \frac{1}{ \sqrt{s_u} }  \frac{ dW(kR) } { ds_0(R) } \, \delta_{\rm m}( \bm{k}  ),  
\eeq
where 
\beq
 s_u(R) =  \int \frac{dk}{  2\pi^2 }  k^2 P_{\rm m}(k)  \bigg(  \frac{dW(kR)}{d s_0(R)} \bigg)^2 . 
 \eeq
In real space, these variables are 
\begin{align}
  \nu ( \bm{r} ) &= \frac{ 1 }{ \sqrt{s_0} } \int d^3r' W( r'; R ) \delta_{\rm m} (\bm{r} -  \bm{r}' ) = \frac{ \delta( \bm{r}) }{ \sqrt{s_0}  } ,  \\
  x ( \bm{r} ) &= - \frac{ 1 }{ \sqrt{s_2} } \int d^3r' \nabla_{r'}^2 W( r'; R )  \delta_{\rm m} (\bm{r} -  \bm{r}' ) = - \frac{ \nabla^2_r  \delta( \bm{r}) }{ \sqrt{s_2} }    ,  \\
  u ( \bm{r} ) &=  \frac{ 1 }{ \sqrt{s_u} } \int d^3r' \frac{ d W( r'; R )}{d s_0(R) }  \delta_{\rm m} (\bm{r} -  \bm{r}' ) =  \frac{1}{\sqrt{s_u}  }   \frac{ d \delta( \bm{r})  }{  d s_0(R) } ,
\end{align}
where we have denoted the smoothed dark matter density field by $ \delta $.  These variables illustrate a few of the ways in which the smoothing window $W$ appears in the constrained variables.  We are now ready to consider generic relations between the constraints.  

The constrained Gaussian field is still Gaussian and its mean and variance are well-known \cite[see e.g. Appendix~D in][]{BBKS86}.  In particular when $C$ is constrained to have some specific values $\mathcal{C} $, then the conditional mean of $\Delta $  is 
\beq
\label{eq:Gaussian_constrained_mean}
\langle \Delta | C = \mathcal{C} \rangle =  \langle \Delta C  \rangle_j  \langle C C  \rangle^{-1}_{\quad j k }  \mathcal{C}_{k}  , 
\eeq
where the vector  $\langle \Delta C  \rangle $ denotes the cross correlation between $\Delta $ and the constraint variables, and  $ \langle C C \rangle $ is the covariance matrix between the constraint variables. 

Clearly, the right hand side of Eq.~\ref{eq:Gaussian_constrained_mean} is a sum of many terms.  If we multiply  $ \langle \Delta  C \rangle  \langle  C C \rangle^{-1}  $ first, then the terms will be grouped according to $\mathcal{C}_{k}$.   For bias, we instead wish to group terms by their scale dependence, which means we group by the elements of $\langle \Delta  C  \rangle $.  We define  the linear bias coefficients as
\beq
\label{eq:DeltaConstrained_lin_bias_def_general}
 \langle \Delta |C =  \mathcal{C}  \rangle = \frac{ \langle  \Delta C \rangle_j  }{ \langle  C C \rangle_{ 1 j } } \sqrt{s_0}   b_1^{(j)}, 
\eeq
where $b_1^{(j)} $ is the $j^{\rm th} $ linear bias coefficient
\beq
\label{eq:lin_bias_def_general}
b_1^{(j)} = \frac{ 1}{\sqrt{s_0} }  \langle C C \rangle_{1 j}  \langle  C C  \rangle^{-1}_{\quad j k }  \mathcal{C}_{k} .  
\eeq
(Note that $j$ is not summed over in Eq.~\ref{eq:lin_bias_def_general}.)  In Eq.~\ref{eq:DeltaConstrained_lin_bias_def_general} we have divided by  $\langle  C C  \rangle_{ 1 j } $  and introduced $\sqrt{s_0}  $ so that the dimension of the resultant bias parameter  $b_1^{(j)} $ agrees with the usual expansion in the density contrast. Note that  $\sqrt{s_0} $ appears quite differently from other variables, so we shall treat it differently (e.g., when we consider bias at different times).  Sometimes it is advantageous to use the variable  $\sqrt{s_0} b_1^{(j)} $ because it can be completely expressed in terms of the normalized dimensionless variables. We also stress that the $b_1^{(j)}$ defined here have not (yet) been averaged over the constraint, i.e.~they explicitly depend on $ \mathcal{C}$.


Eq.~\ref{eq:lin_bias_def_general} is interesting because it enables us to express the bias parameters directly in terms of the constraints  $\mathcal{C} $, making the bias problem simply a linear algebra problem. In particular, we can invert Eq.~\ref{eq:lin_bias_def_general} to express $\mathcal{C}_{k} $ in terms of the $b_1^{(j)}$: 
\beq
\label{eq:consistency_relation}
\mathcal{C}_{k} = \langle C C \rangle_{k j }  \frac{\sqrt{s_0} \,  b_1^{(j)} }{\langle C C \rangle_{1j}  }. 
\eeq
There are $n$ relations between the bias parameters, where $n$ is the dimension of $\langle CC\rangle$, and they simply reflect the underlying constraints on halo formation.  Note that the bias in Eq.~\ref{eq:lin_bias_def_general} and the consistency relations in Eq.~\ref{eq:consistency_relation} only depend on $\mathcal{C}$, but not $\Delta $. This echoes our previous assertion that the large-scale field $\Delta $ is only used to extract the constraint $\mathcal{C}$. 

In addition, recall that we have yet to average over the constraint variables.  Since averaging is a linear operation, our analysis also shows that partial averages will also satisfy similar consistency relations.  As we discuss later, this means that our analysis is immune to what is known as `assembly bias'.  I.e., although the numerical values of `assembly biased' bias factors may be different from those of the parent population, the algebraic relations between the bias factors -- Eq.~\ref{eq:consistency_relation} -- will be the same as for the parent population.  

The full consistency relations are already given by Eq.~\ref{eq:consistency_relation}. However,  the  first of these consistency relations for $\mathcal{C}_1$ is particularly simple. By putting $k=1$ in Eq.~\ref{eq:consistency_relation}, we get
\beq
\label{eq:consistency_relation_first}
\sum_j b_1^{(j)} = \frac{ \mathcal{C}_1} {\sqrt{s_0} }. 
\eeq


Eq.~\ref{eq:consistency_relation_first} is interesting because it shows that the value of the constrained variable $C_1$ can be determined if one has measured the bias parameters $b_1^{(j)}$.  This shows explicitly why measurements of the large scale bias can be used to learn about the small-scale physics of halo formation.  Notice in particular that if the constrained variables must be averaged over, then this will yield averaged values of $b_1^{(j)}$ whose values will depend on the range over which the average was taken.  Inserting these averaged values in Eq.~\ref{eq:consistency_relation_first} yields the corresponding averaged value of $C_1$.  Since the same procedure works whatever the range over which the average was taken, Eq.~\ref{eq:consistency_relation_first} shows that the bias factors of an `assembly biased' population can be used to infer what was the physics which led to the assembly bias.  While one might have worried that the procedure for recovering the physics of formation might have been different for each sub-population, the analysis above shows that this is not the case.  

Eq.~\ref{eq:consistency_relation_first} has been derived previously, in the context of a specific model in which $n=2$, following a rather different approach \citep{MussoParanjapeSheth2012,ParanjapeShethDesjacques2013}, where it was called a `consistency' relation.  Our analysis shows how to extend the consistency relation to arbitrary $n$ and arbitrary constraints.  On the other hand, the analysis above is for the first order (sometimes called `linear') bias parameter.  There are consistency relations associated with higher order bias parameters \citep{MussoParanjapeSheth2012,ParanjapeShethDesjacques2013}.  Based on the results of \cite{MussoParanjapeSheth2012}, we write down the  high order consistency relations in Sec.~\ref{sec:revisit}.   \cite{CastorinaSheth2017} show how to extend these to $n$ arbitrary variables and constraints using a generating function approach.  So it would be interesting to extend the simple linear algebra method developed here to describe higher order bias.  However, doing so is beyond the scope of this work.

\subsection{Smoothing windows}\label{sec:Ws}
The analysis above is valid regardless of the form of $W$.  We now turn to the question of what one should use for $W$.  We expect it to have a characteristic scale $R$, so its Fourier transform will be a function of the combination $kR$.  In what follows, we will use the effective window function proposed in \cite{ChanShethScoccimarro2015}, the Fourier transform of which is 
\begin{align}
 \label{eq:Weff}
 W_{\rm Eff}(kR) &= W_{\rm TH}(kR)\, W_{\rm G}(kR/5), 
\end{align}
where  $W_{\rm TH}$ and $W_{\rm G}$ denote the top hat and Gaussian window function respectively:
\begin{align}
\label{eq:Wth}
 W_{\rm TH} (x) &=  (3/x^3)\, ( \sin x - x \cos x ),   \\ 
 W_{\rm G} (x) &=   e^{-x^2/2}. 
\end{align}

There are a number of reasons why we will work with $W_{\rm Eff}$, rather than $W_{\rm TH}$ or $W_{\rm G}$.  But before we list them, it is worth noting that although all three filters tend to unity at $x\ll 1$, they are not all equally compact.  A crude measure of the extent of these filters is given by expanding $W$ to lowest order in $k$.  Then
 $W_{\rm Eff}(x)\approx 1 - 3x^2/25$, 
 $W_{\rm TH}(x)\approx 1 - x^2/10$, and 
 $W_{\rm G}(x)\approx 1 - x^2/2$, 
which suggests that, if we wanted to replace $W_{\rm Eff}$ by a tophat or a Gaussian, then we should set $R_{\rm G}^2= (6/25) \, R^2_{\rm Eff}$ and $R_{\rm TH}^2 = (6/5)\,R_{\rm Eff}^2$.  Thus, the tophat is the most compact of the three, and the Gaussian is the most extended.  This difference will be important below.  

The first reason we like $W_{\rm Eff}$ is purely technical.  Although many of the expressions above are formally well-defined for an arbitrary smoothing window, in practice, some lead to divergences.  For example, $s_j$ defined in Eq.~\ref{eq:sj} plays an important role in peak theory.  However, for $W_{\rm TH}$, the integral above diverges if $j\ge 2$, whereas it is well-behaved for $W_{\rm Eff}$.

The second is more physical.  With $W_{\rm Eff}$, \cite{ChanShethScoccimarro2015} were able to provide an accurate description of protohalos in simulations using just two bias parameters:  those associated with $\nu$ and $u$.  In contrast, with $W_{\rm G}$, just $\nu$ and $u$ are not enough to accurately describe protohalos.  That is to say, $W_{\rm Eff}$ allows us to use the $n=2$ consistency relations above to constrain halo formation.  I.e., of the potentially vast number of physical parameters which may matter for halo formation, $W_{\rm Eff}$ allows us to work with just $n=2$.  This is a very useful simplification.  

Finally, $W_{\rm Eff}$ allows us to illustrate some subtle issues associated with our approach.  For example, $x=u$ for a Gaussian smoothing window because 
\beq
\label{eq:u_to_x_Gaussian}
  \frac{dW_{ \rm G }(kR)}{d\ln s_0(R)} 
 =  \frac{ s_0(R) }{2 R^2s_1(R) } \, k^2R^2\, W_{\rm G}( kR ) .
\eeq
Therefore, for general smoothing windows, such as our Eq.~\ref{eq:Weff}, we expect to find that although $x$ and $u$ are formally different, in practice, constraints on $x$ may be rather degenerate with those on $u$.  

\begin{figure}
\centering
\includegraphics[width=\linewidth]{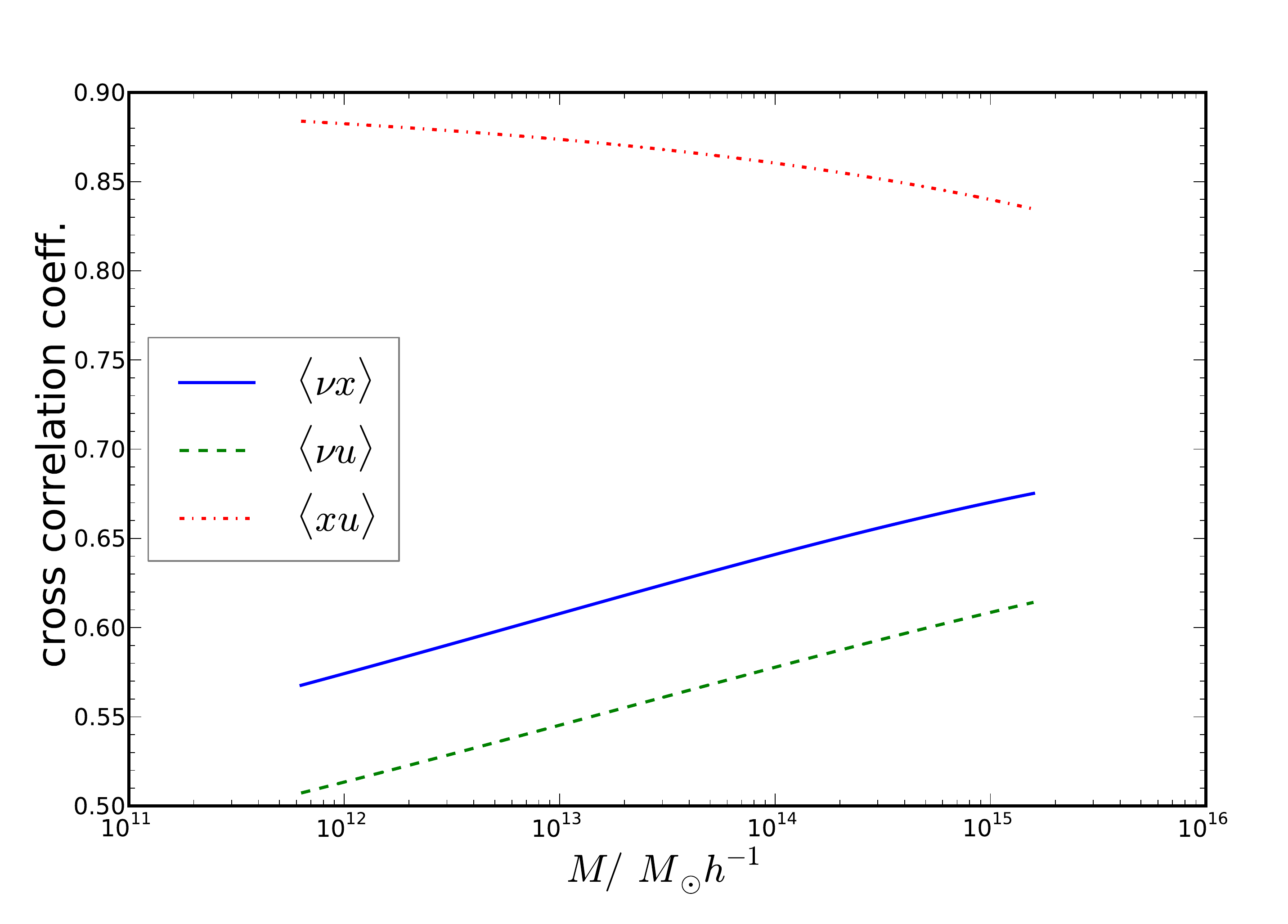}
\caption{ Cross correlation coefficients   $\langle \nu x  \rangle$ (solid blue), $\langle \nu u \rangle  $ (dashed green) and  $\langle xu \rangle  $ (dotted-dashed red) as a function of halo mass, when the smoothing window is given by Eq.~\ref{eq:Weff}.  }
\label{fig:cross_coef}
\end{figure}

To illustrate this point, Fig.~\ref{fig:cross_coef} shows the cross correlation coefficients
\begin{align}
\label{eq:nu_x_cc}
 \langle \nu   x \rangle  &= \frac{s_1  }{\sqrt{s_0 s_2 }} \equiv  \gamma_{\nu x} ,   \\
\label{eq:nu_u_cc}
\langle \nu  u  \rangle  &= \frac{ 1/2 }{ \sqrt{s_0 s_u }   }\equiv  \gamma_{\nu u} ,  \\
\label{eq:x_u_cc}
\langle  x u  \rangle  &= \frac{ 1/2 }{ \sqrt{s_2 s_u } } \frac{ d s_1 }{ d s_0 } \equiv \gamma_{xu}. 
\end{align}
as a function of smoothing scale $R$ (recall halo mass $M\propto R^3$).  Notice that $\langle xu\rangle \approx 1$, and $\langle \nu x \rangle $ and $\langle \nu  u \rangle $ are  similar in magnitude, as expected (recall that $\langle x  u \rangle = 1$ for Gaussian smoothing).  In what follows, we will use this fact to address a number of issues which arise when different physical parameters are nearly degenerate.

\subsection{ Specific examples }
\label{sec:specific}

It is common to set $C_1 = \nu$, although this is not necessary.  In the simplest peak model, a proto-halo patch satisfies three constraints:  the value of the enclosed overdensity (the height),  the first derivative with respect to spatial position (which must be zero to be a local peak), and  the second derivative with respect to spatial position (the peak curvature) \citep{BBKS86}.   In the simplest excursion set peak model, there is an additional constraint on the first derivative with respect to smoothing scale (the excursion set slope) \citep{AppelJones1990, MussoSheth2012, ParanjapeSheth2012}.  More sophisticated models include the shear and the shape  \citep[e.g.][]{ShethChanScoccimarro2013, BiagettiChanetal_2014,CastorinaParanjapeHahnSheth2016}; we will ignore these complications.

The first spatial derivative is independent of the other three variables, and it is constrained to be zero anyway.  So, in practice, $n=2$ for the peak model and  $n=3$ for excursion set peak.  It is typical to order these as $ C = (\nu,x,u)$, where the three variables have zero mean, and they have been normalized by the rms values of the height, curvature and slope, respectively.  Therefore, $\langle C  C \rangle $ has unity along its diagonal.  In addition, since $x\approx u$, we will use this example to show how our formulation of consistency relations works when $n\ge 2$, as well as to address the question of what one should do when different physical parameters are nearly degenerate.  

E.g., if $x$ and $u$ are nearly degenerate, shouldn't the bias factors for each be similar?  If so, how is it that the $n=3$ consistency relation does not double count the contribution from the two terms?   Alternatively, what form does the consistency relation take if fundamentally both $x$ and $u$ matter, but we suppose that only one of them matters?  Can we effectively group the $b_1^{(2)}$ and  $b_1^{(3)} $ terms together in the consistency relations?  Such questions are relevant to the issue of efficiency in describing the halo distribution:  How many variables are sufficient to capture the physics?  In what follows, we will consider both the 2-parameter and 3-parameter models. 

The previous section showed that linear combinations of the bias factors allow one to isolate the dependence of halo formation on $\nu$, $x$ and $u$.  To see this in practice, suppose that  $n=1$ and $C_1=\nu$. Then
\beq 
\label{eq:b1_n1}
 b_1^{(1)} = \nu_{\rm c}/\sqrt{s_0} ,
\eeq
where $\nu_{\rm c}$ would typically be given by the spherical collapse (SC) value $\nu_{\rm sc}$
\beq
\label{eq:nu_c}
 \nu_{\rm sc} = \delta_{\rm sc}(z)/\sqrt{s_0} ,
\eeq
with 
\beq
\label{eq:deltac_z}
 \delta_{\rm sc}(z) = \frac{  D(z=0) }{   D(z)  } \delta_{\rm sc},  
\eeq
where $\delta_{\rm sc} =1.68$ is the SC threshold and $D$ is the linear growth factor.  
Eq.~\ref{eq:b1_n1} is the expression for biased tracers which appears in \cite{Kaiser84}.  If the biased tracers were constrained to have $\delta\ge\delta_{\rm sc}$ rather than $\delta=\delta_{\rm sc}$, then, because the constraint $C$ would include an integral over $\delta\ge\delta_{\rm sc}$, the bias factor $b_1^{(1)}$ would also be given by integrating over the allowed range, and $\nu_{\rm sc}$ would also be replaced by its average over the range. Note that for notational simplicity we have denoted the smoothed dark matter density $W*\delta_{\rm m} $ simply as $\delta$, and the window function is implicitly included in $\delta$ and the constraint variables.   

The next non-trivial case is $n=2$, with $C =(\nu,x)$ in the peak model, and $( \nu, u )$ in the excursion set approach, where $\nu$ is the overdensity as before, $u$ its slope with respect to smoothing scale, and $x$ its curvature with respect to spatial position.  E.g., in the peak model, 
\begin{equation}
\label{eq:CC_2}
 \langle CC \rangle  =  \left( \begin{array}{cc}
                         1       &  \gamma_{\nu x}  \\
                         \gamma_{\nu x}  &     1    \end{array}   \right) , 
\end{equation}
and the bias expansion is given by 
\beq
\langle \Delta | C=( \nu_{\rm c}, x_{\rm c}) \rangle  = \langle \Delta \delta \rangle b_{1}^{(1)}  -   \langle \Delta \nabla^2 \delta \rangle \frac{ s_0 }{ s_1 } b_1^{(2)}, 
\eeq
with the bias parameters being 
\begin{align}
 \label{eq:bnux1}
 b_{1}^{(1)} &= \frac{\nu_{\rm c} - \gamma_{\nu x}  x_{\rm c}  }{ \sqrt{s_0} ( 1 - \gamma_{\nu x}^2 ) } \\
 \label{eq:bnux2}
 b_{1}^{(2)} &=  \frac{ \gamma_{\nu x}  ( x_{\rm c} - \gamma_{\nu x}\nu_{\rm c}  ) }{ \sqrt{s_0} ( 1 - \gamma_{\nu x}^2 ) } ,
\end{align}
where $\nu_{\rm c}$, and  $ x_{\rm c} $ denote the constrained values.  The consistency relations read
\begin{equation}
\label{eq:ConsistencyRelation_2}
 \left( \begin{array}{c}     
\nu_{\rm c}    \\
x_{\rm c}       \end{array}   \right) 
                         = \sqrt{s_0}   \left( \begin{array}{c}
                                               b_{1}^{(1)} + b_{1}^{(2)}   \\
                         \gamma_{\nu x} b_{1}^{(1)} + b_{1}^{(2)}/\gamma_{\nu x}  \end{array}   \right). 
\end{equation}
For the excursion set approach, simply replace $\gamma_{\nu x}\to \gamma_{\nu u}$ and $ - \langle \Delta \nabla^2 \delta \rangle (s_0/s_1) \to2  \langle \Delta  d \delta /d\ln s_0\rangle$.  

Notice that $b_1^{(1)}+b_1^{(2)} = \nu_c/\sqrt{ s_0} $, just as when $n=1$.  However, $b_1^{(1)}$ here will be different from when $n=1$ if $b_1^{(2)}\ne 0$.  Again, allowing a range of $\delta_c$ values means one must replace $b_1^{(1)}$ and $b_1^{(2)}$ by their averaged values, and $\nu_c/\sqrt{s_0}$ should also be replaced by its average.  

E.g., a simple model of assembly bias would be to assume that all objects have the same height $\delta_c$, but different curvatures or slopes \citep{Dalaletal2008, MussoSheth2014,Lazeyras:2016xfh}.  Eq.~\ref{eq:bnux1} and \ref{eq:bnux2} show that populations with different $x_c$ will have different bias parameters $b_1^{(1)}$ and $b_1^{(2)}$.  However, because we have assumed they have the same height $\nu_c$, the sum $b_1^{(1)}+b_1^{(2)}$ (the first consistency relation) will be the same for the different populations.  On the other hand, because the left hand side of the second consistency relation equals $x_c$, this one will be different for the different populations.  Thus, measuring the large scale bias factors $b_1^{(1)}$ and $b_1^{(2)}$ for the populations and then using the two consistency relations in Eq.~\ref{eq:ConsistencyRelation_2} allows one to discover that the populations only differ in curvature and not in height.  Since this argument obviously holds if the different populations are constrained to have different ranges in $\delta$ and $x$, this example shows explicitly how the same methodology can be used to learn about the physics of assembly bias, without having a priori knowledge that it was even present in the first place. Conversely, if we have some way to directly measure $\delta_{\rm c}$ and $x_{\rm c}$, e.g.~using the direct measurements outlined in Sec.~\ref{sec:cross_correaltion_CR},  then the consistency relations provide a way to detect assembly bias. 

Finally, we consider $n=3$ with $ C = (\nu,x,u)$; recall that $u$ is associated with the excursion set upcrossing constraint on the slope of the smoothed density field (Eq.~\ref{eq:u_def}).  The bias expansion reads
\begin{align}
\label{eq:bias_realspace_3}
\langle \Delta | C=( \nu_{\rm c}, x_{\rm c}, u_{\rm c}) \rangle & = \langle \Delta \delta \rangle b_{1}^{(1)}  -   \langle \Delta \nabla^2 \delta \rangle \frac{ s_0 }{ s_1 } b_{1}^{(2)}  \nn \\
     & +\quad  2  \Big\langle  \Delta  \frac{d \delta  }{d s_0 } \Big\rangle s_0 b_1^{(3)} , 
\end{align}
and Eq.~\ref{eq:consistency_relation} becomes
\begin{equation}
\label{eq:ConsistencyRelation_3}
 \left( \begin{array}{c}     
 \nu_{\rm c}    \\ 
   x_{\rm c}    \\  
   u_{\rm c} \end{array} \right) 
                         = \sqrt{s_0}   \left( \begin{array}{c}
                             b_{1}^{(1)} + b_{1}^{(2)} +  b_{1}^{(3)}   \\
              \gamma_{\nu x} b_{1}^{(1)} + b_1^{(2)}/\gamma_{\nu x} +  b_1^{(3)}\,\gamma_{x u}/\gamma_{\nu u} \\ 
              \gamma_{\nu u} b_{1}^{(1)} + b_1^{(2)}\,\gamma_{x u}/\gamma_{\nu x}   + b_1^{(3)}/\gamma_{\nu u}  
      \end{array}    \right). 
\end{equation}
In this form it is easy to see that the consistency relation is satisfied.  Clearly, our previous remarks about assembly bias apply here too.  Our main goal in this example is to illustrate what happens when some of the parameters, in this case $x$ and $u$, are nearly degenerate.  
E.g., for Gaussian smoothing $\langle xu\rangle = 1$, and $\langle\nu u\rangle = \langle\nu x\rangle$ so the expressions above clearly depend only on $b_1^{(1)}$ and $(b_1^{(2)} + b_1^{(3)})$, and the relations for $x_c$ and $u_c$ are the same.  

The bias factors, written explicitly in terms of the correlation coefficients of the constrained variables, are
\begin{align}
 \label{eq:bnuxu}
 b_{1}^{(1)} &= \frac{\nu_{\rm c} - \gamma_{\nu x}\, x_{\rm c}  }{ \sqrt{s_0} ( 1 - \gamma_{\nu x}^2 ) } 
              - \frac{\gamma_{\nu u} - \gamma_{u x}\gamma_{x\nu}}{1-\gamma^2_{\nu x}}\,
                \frac{b_1^{(3)}}{\gamma_{\nu u}},   \\
 b_{1}^{(2)} &=  \frac{ \gamma_{\nu x}  ( x_{\rm c} - \gamma_{\nu x}\,\nu_{\rm c}  ) }
                   { \sqrt{s_0} ( 1 - \gamma_{\nu x}^2 ) }
             - \gamma_{\nu x} \frac{\gamma_{ux} - \gamma_{u\nu}\gamma_{\nu x}}{1-\gamma^2_{\nu x}}
                \frac{b_1^{(3)}}{\gamma_{\nu u}},   \\  
 b_{1}^{(3)} &= \frac{ \gamma_{\nu u}}{\sqrt{s_0} }  \,\frac{u_c - \langle u|\nu_c,x_c\rangle}{{\rm var}(u|\nu_c,x_c)},
\end{align}
where 
\beq
 \langle u|\nu_{\rm c},x_{\rm c} \rangle = \frac{ ( \gamma_{\nu u} - \gamma_{ xu} \gamma_{ \nu x} ) \nu_{\rm c} + (\gamma_{ x u } - \gamma_{\nu x} \gamma_{\nu u} ) x_{\rm c}   }{  1 - \gamma_{\nu x}^2   }
\eeq
and 
\beq
\label{eq:var_u_nux}
 {\rm var}(u|\nu_{\rm c},x_{\rm c}) = \frac{1 - \gamma_{\nu x}^2 - \gamma_{ux}^2 - \gamma_{u\nu}^2 + 2\gamma_{\nu x}\gamma_{\nu u}\gamma_{ux}}{1 - \gamma_{\nu x}^2}.
\eeq
The first terms on the rhs of the expressions for $b_1^{(1)}$ and $b_1^{(2)}$ are the same as when $n=2$ (compare Eq.~\ref{eq:bnux1} and \ref{eq:bnux2}), so it appears to be straightforward to address the question of the impact of the third variable. However, when $\gamma_{ x u } \rightarrow 1  $, both  $ ( u_c - \langle u|\nu_c,x_c\rangle ) $ and  var$(u|\nu_{\rm c},x_{\rm c} ) $ vanish, thus  $b_1^{(3)} $ and hence $b_1^{(1)} $ and $ b_1^{(2)}$ are indeterminate. As we discussed,  in this limit, the problem indeed reduces to the $n=2$ case, and we can choose the second variable to be  either $x$ or $u$. We will return to this shortly.

\begin{figure}
\centering
\includegraphics[width=\linewidth]{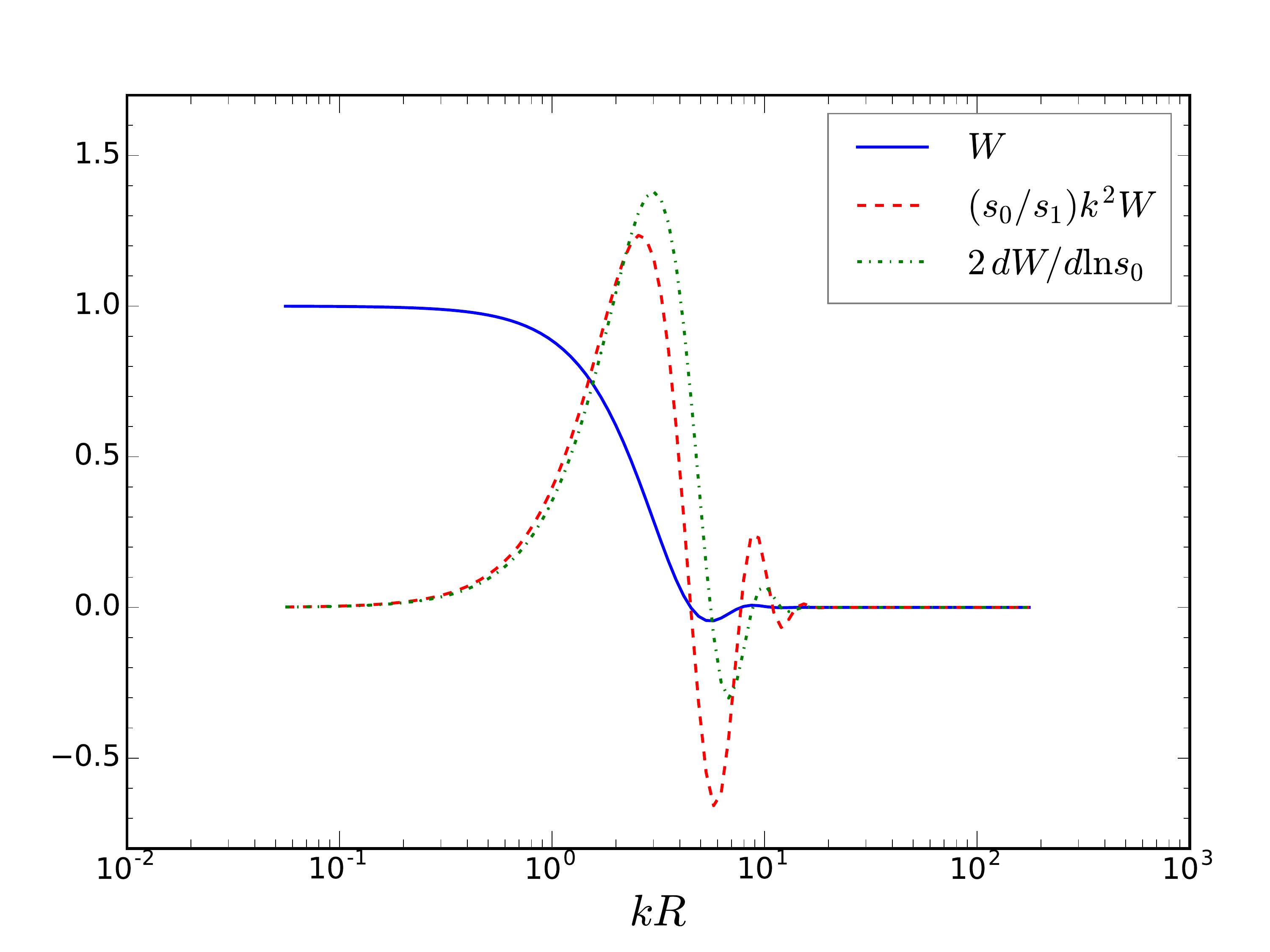}
\caption{ The three terms contributing to the effective cross bias in Eq.~\ref{eq:bceff_3}, when $M=5 \times 10^{13} \Msun$ and the smoothing window is given by Eq.~\ref{eq:Weff}. The dashed and dot-dashed terms, which are due to the curvature and slope constraints, have rather similar  shapes. }
\label{fig:deg_terms_compare}
\end{figure}

We now turn to the question of estimating the bias coefficients from cross-correlations with the dark matter density field $\Delta$.  The scale dependence of bias comes from the interplay of three terms:  $\langle\Delta\nu\rangle$, $\langle\Delta x\rangle$ and $\langle\Delta u\rangle$, which are defined by
\begin{align}
 \langle \Delta \nu  \rangle
 & =\frac{1}{\sqrt{ s_0} }  \int \frac{dk}{2\pi^2  } \, k^2 P_{\rm m} (k) \, W(kR)  ,    \\
  \langle \Delta x \rangle 
 & = \frac{ 1 }{ \sqrt{ s_2}  } \int \frac{dk}{ 2\pi^2  } \,  k^4 P_{\rm m}(k)  W(kR) ,   \\
 \langle\Delta u \rangle  
 & = \frac{ 1 }{ \sqrt{ s_u } }  \int \frac{dk}{ 2 \pi^2  }\,  k^2P_{\rm m}(k) \,
                       \frac{dW(kR)}{d s_0(R)} .
\end{align}
(We note in passing that in these expressions there is a window function only for the protohalo/peak; there is no window on $\Delta$.  Also, for a Gaussian window, $ \langle \Delta u \rangle = \langle\Delta x \rangle$ so we expect this to remain approximately true for other windows.)  Fourier transforming Eq.~\ref{eq:bias_realspace_3}, and dividing by the cross correlation between $\Delta $ and the dark matter density contrast $\delta_{\rm m }$, yields
\begin{align}
\label{eq:bceff_3} 
b_{\rm eff}^{(3)} ( k )   &\equiv  \frac{ \langle \Delta | C = \mathcal{C}  \rangle   }{ \langle  \Delta \delta_{\rm m}  \rangle   }  \nn \\
&=   W  b_{1}^{(1)}   +  \frac{s_0}{R^2 s_1}\, (kR)^2\, W   b_{1}^{(2) }   + 2  \frac{  d W }{d \ln s_0 }  b_{1}^{(3)}  .
\end{align}
Fig.~\ref{fig:deg_terms_compare} compares the scale dependence of the three terms on the rhs when $W$ is given by Eq.~\ref{eq:Weff}.  For $kR \lesssim 0.5$, only the first term is non-vanishing:  hence, $b_1^{(1)}$ can be determined from the bias signal at $kR \lesssim 0.5$.  The other two terms are rather similar.  Therefore, when Eq.~\ref{eq:bceff_3} is used to fit the $k$-dependence of the measured bias signal, the derived constraints on $b_1^{(2)}$ and $b_1^{(3)}$ are likely to be degenerate, along a locus of approximately constant $b_1^{(2)} + b_1^{(3)}$.  

Having shown why $b_1^{(2)}$ and $b_1^{(3)}$ are approximately degenerate, we are ready to reconsider the issue of var$(u|\nu,x)\to 0$.  It is particularly instructive to consider the following variables 
\beq
 x_\pm \equiv \frac{x \pm u}{\sqrt{2(1\pm\gamma_{xu})}}
 \quad{\rm for\ which}\quad 
 \gamma_{\nu\pm} \equiv \frac{\gamma_{\nu x}\pm\gamma_{\nu u}}{\sqrt{2(1\pm\gamma_{xu})}}.
\eeq  
These variables are uncorrelated with each other, and this simplifies the analysis.  By analogy with Eq.~\ref{eq:bnuxu}-\ref{eq:var_u_nux}, the bias factors are 
\begin{align}
 b_1^{(1)} &= \frac{\nu_{\rm c} - \gamma_{\nu +}\, x_{+c}  }{ \sqrt{s_0} ( 1 - \gamma_{\nu +}^2 ) } 
              - \frac{b_1^{(-)}} {1-\gamma^2_{\nu +}}\,,   \\
 b_1^{(+)} &=  \frac{ \gamma_{\nu +}  ( x_{+c} - \gamma_{\nu +}\,\nu_{\rm c}  ) }
                   { \sqrt{s_0} ( 1 - \gamma_{\nu +}^2 ) }
             + \frac{\gamma_{\nu +}^2 b_1^{(-)} }{1-\gamma^2_{\nu +}},   \\  
 b_1^{(-)} &=  \frac{\gamma_{\nu -}}{ \sqrt{s_0}  }  \,\frac{x_{-c} - \langle x_-|\nu_c,x_{+c} \rangle}{{\rm var}(x_-|\nu_c,x_{+c})}
\end{align}
where 
\beq
 \langle x_-|\nu_c,x_{+c}\rangle =  
    \frac{\gamma_{\nu -}}{1-\gamma_{\nu +}^2}(\nu_c - \gamma_{\nu +} x_{+c})
\eeq
and 
\beq
 {\rm var}(x_-|\nu_c,x_{+c}) = \frac{1 - \gamma_{\nu +}^2 - \gamma_{\nu -}^2}{1 - \gamma_{\nu +}^2}.
\eeq
It is again easy to see that the consistency relation is the same as before.  However, now var$(x_-|\nu_c,x_{+c})$ is well-behaved, so it is clear that, when $\gamma_{\nu -} \ll 1$ (when $x =u$, $\gamma_{\nu -} $ is still indeterminate), then $b_1^{(-)}\ll 1$ and the system reduces to one involving just two constrained variables, $\nu_c$ and $x_{+c} = (x_c+u_c)/2$.

\begin{figure}
 \centering
 \includegraphics[width=\linewidth]{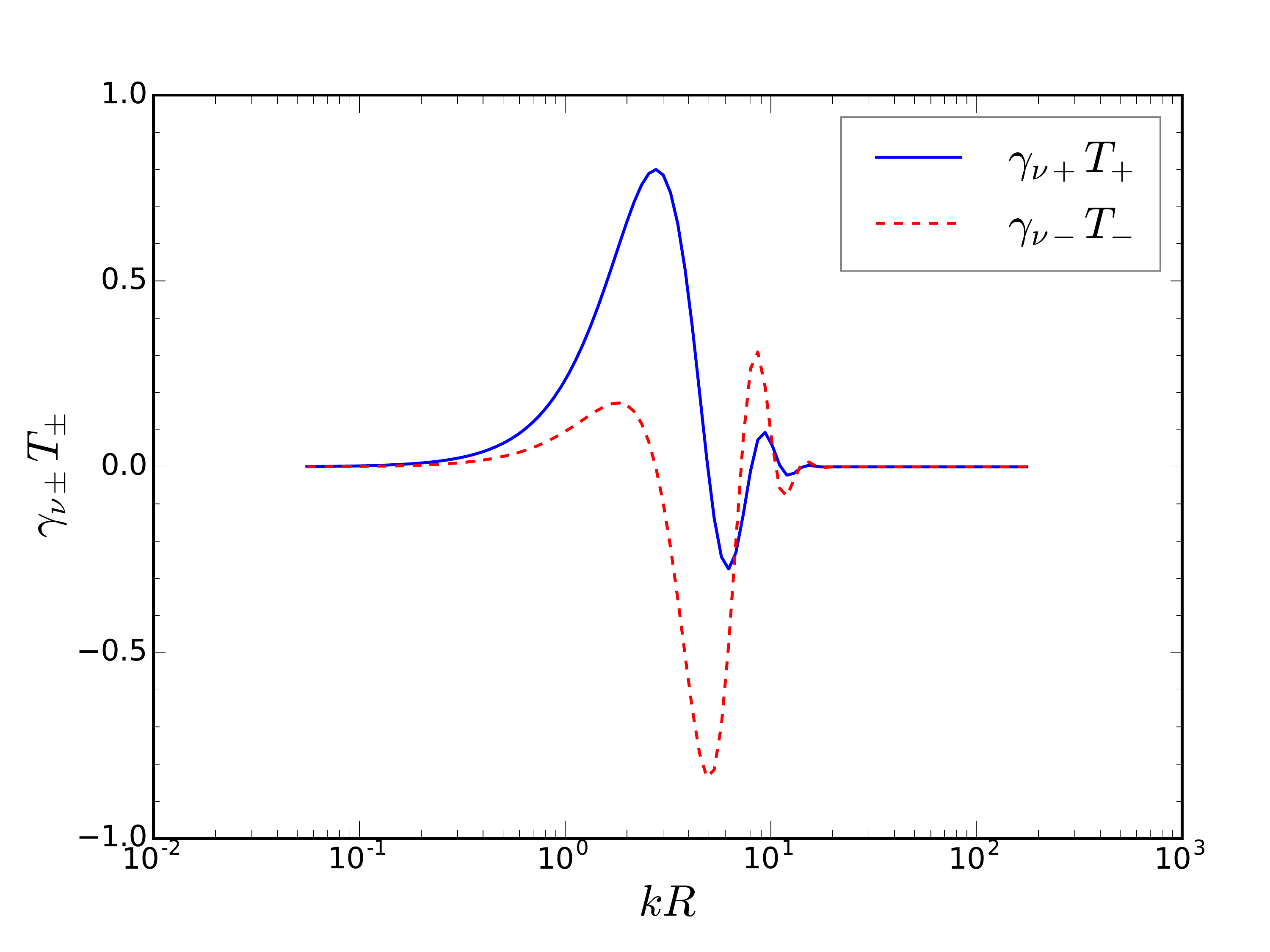}
 \caption{The terms  $ \gamma_{\nu + } T_{+}$ (solid blue) and $ \gamma_{\nu -}   T_{-}$ (dashed red)  when $M=5 \times 10^{13} \Msun$, shown as a function of $kR$.  Note that $ \gamma_{\nu -} T_-$ remains close to zero compared to $ \gamma_{\nu + } T_{+}$ until  $kR \sim 3 $.    }
 \label{fig:Tplus_min}
\end{figure}

Using Eq.~\ref{eq:DeltaConstrained_lin_bias_def_general}, the corresponding effective bias expansion reads
\beq
\label{eq:bef3_Tpm}
b_{\rm eff}^{ 3'}  =  b_1^{(1)}  W +   b_1^{(+)}  T_{+}   +  b_1^{(-)}  T_{-} ,  
\eeq
where
\beq
 T_\pm =  \frac{ s_0 k^2  }{ s_1 } W  \frac{\gamma_{\nu x}} {\gamma_{\nu x} \pm\gamma_{\nu u}} 
            \pm 2 \frac{dW}{d\ln s_0} \frac{\gamma_{\nu u}} {\gamma_{\nu x} \pm\gamma_{\nu u}} .
\eeq
(In the $\pm$ sign, $+$ and $-$ are for $T_+$ and $T_-$ respectively.)  

The quantities which actually matter for the bias expansion are $\gamma_{\nu\pm}T_{\pm}$.  Fig.~\ref{fig:Tplus_min} shows that for $kR\lesssim 3$, $\gamma_{\nu -}T_{-}\ll \gamma_{\nu +}T_{+}$ and so the $b_1^{(-)}T_{-}$ term can be neglected.  Therefore, when fitting to data at $kR \lesssim 3$, it should be a good approximation to neglect the $x_{-}$ variable and use only the effective two-parameter model.  This is important, as when fitting data, one does not know a priori how many variables are important for the physics.  The fact that $b_1^{(-)}T_-\ll 1$ means that
working with the effective two-variable problem will not yield biased results.  

\begin{figure*}
\centering
\includegraphics[width=0.9\linewidth]{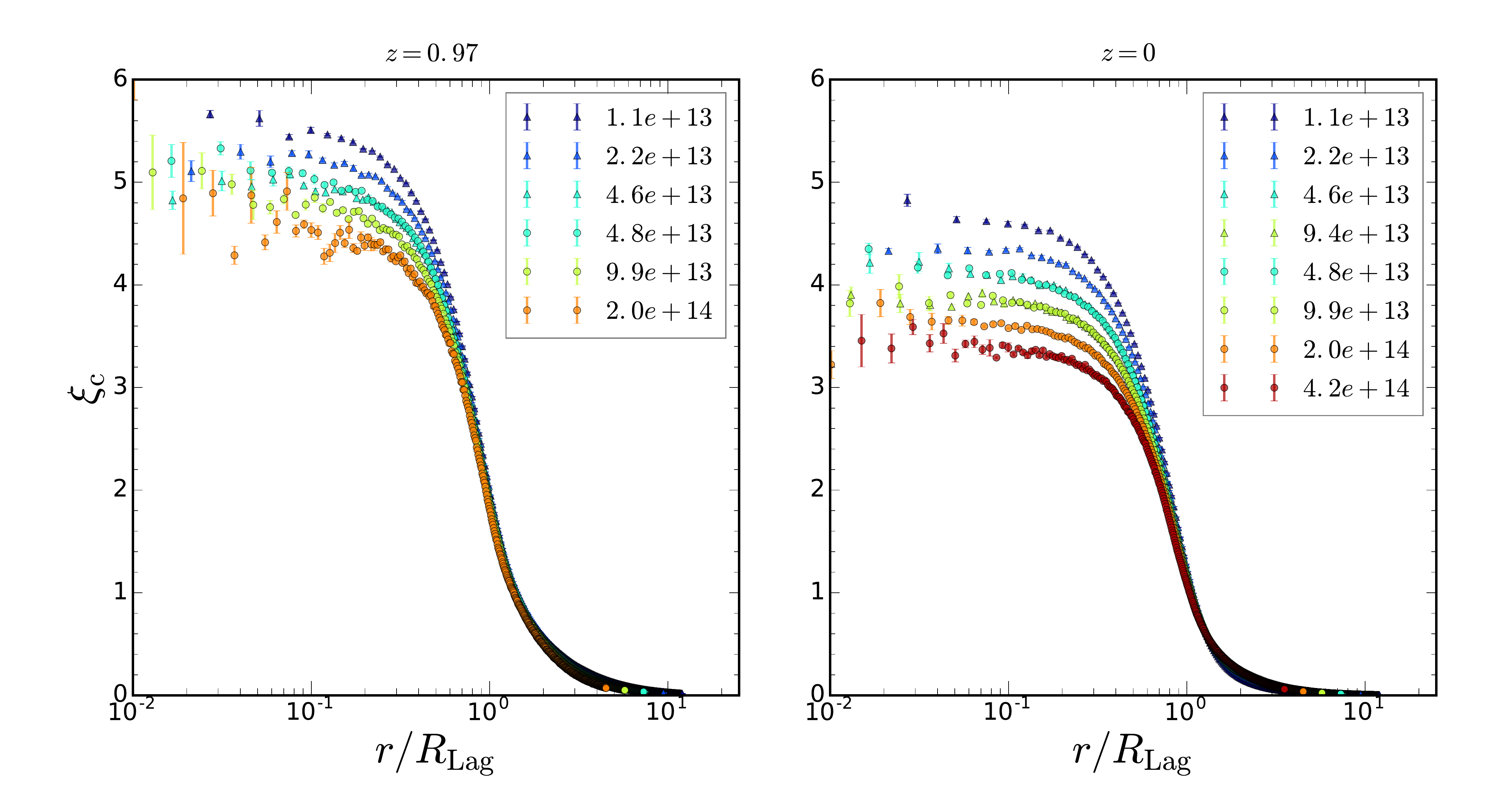}
\caption{Cross correlation between Lagrangian protohalo centers and the dark matter measured in the Oriana (circles) and Carmen (triangles) simulations and extrapolated to  $z=0 $ using the linear growth factor. The corresponding Eulerian halos were identified at $z=0.97$ (left) and 0 (right). Inset lists masses in units of $\Msun $. Less massive protohalos have steeper profiles, in good agreement with a generic prediction of peak theory.}
\label{fig:xic_rnorm}
\end{figure*}

\section{Cross correlations and Lagrangian bias consistency relations}
\label{sec:cross_correaltion_CR}

As discussed in the previous section, the cross correlation between the dark matter field and the Lagrangian protohalos encodes the constraints used to define the Lagrangian protohalos. We shall check this using  Lagrangian protohalos obtained in numerical simulations.  To do so,  we identify Eulerian halos at some redshift $z$, and trace the constituent particles in each halo back to the initial redshift $z_*$. The position of the Lagrangian protohalo is estimated using the center of mass of the particles in the Lagrangian space.

In this work, we use two sets of simulations from the LasDamas project, denoted by Oriana and Carmen respectively. Both sets assume a flat $\Lambda$CDM model with the cosmological parameters, $\Omega_{\rm m} =0.25 $, $\Omega_{\Lambda} =0.75$ and $\sigma_8 =0.8 $.  The initial conditions are Gaussian with spectral index $n_{\rm s}=1$, and transfer function output from \small{CMBFAST} \citep{CMBFAST}.  The initial displacement fields are set using 2LPT \citep{CroccePeublasetal2006} at $z_*=49$. The simulations are evolved using the public code \small{Gadget2} \citep{Gadget2}.  In the Oriana simulations, there are $1280^3$ particles in a cubic box of size 2400 $ \MpcOh $.  For the Carmen simulations, there are $1120^3 $ particles in a box of size 1000 $ \MpcOh $.   Thus the particle mass in Oriana and Carmen is $4.57  \times 10^{11} $ and  $4.94  \times 10^{10}  M_{\odot}h^{-1}$  respectively.  We use 5 realizations from Oriana and 7 from Carmen. In each, the Eulerian halos are identified using a Friends-Of-Friends algorithm \citep{Davisetal1985} with linking length $b=0.156$ at $z=0.97$ and 0.  We only consider halos with at least 65 particles. We bin the halos into thin narrow mass bins of width $\Delta \ln M = 0.157 $. To avoid contamination from evolution, by Lagrangian space we mean the initial Gaussian density field evaluated at the grid positions instead of the displaced 2LPT field.

\subsection{Lagrangian cross-correlations in configuration and Fourier space}
\label{sec:correlation_measurements}
We measure cross-correlations between the initial Lagrangian protohalo positions and the dark matter at the initial time using configuration space and Fourier methods.  

\begin{figure*}
\centering
\includegraphics[width=0.9\linewidth]{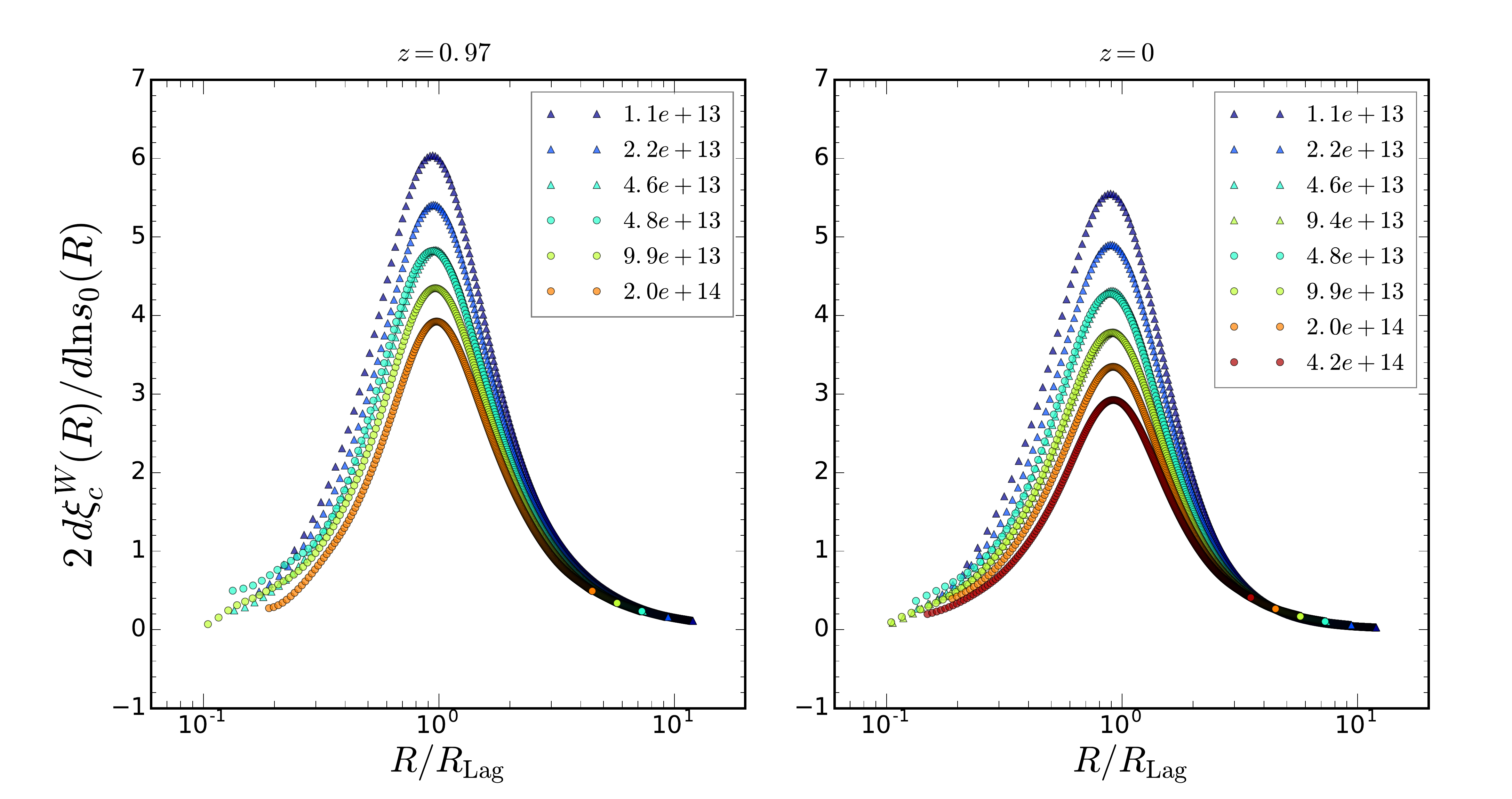}
\caption{ The slope variable $2\,d\xi_c^W/d\ln s_0$, shown as a function of the smoothing scale $R$ expressed in units of $R_{\rm Lag}$.  The slope is maximal around $R_{\rm Lag}$, with less massive halos having steeper slopes.}
\label{fig:u_rnorm}
\end{figure*}

The configuration space signal, $\xi_{\rm c}$ is simply the mean density profile around the Lagrangian halo centre \cite[e.g.][]{Peebles1980}.  Fig.~\ref{fig:xic_rnorm} shows this cross-correlation for a range of narrow mass bins; much of the dependence on mass is removed if distances are scaled by $R_{\rm Lag}$, which is defined as
\beq
 R_{\rm Lag} \equiv \bigg( \frac{ M }{ 4 \pi  \bar{\rho}_{\rm m}/3 }   \bigg)^{1/3} 
\eeq
($\bar{ \rho}_{ \rm m} $ is the comoving density of the dark matter).
However, there is some residual mass dependence: $\xi_{\rm c}$  drops slightly more rapidly for lower mass halos.  The plot shows that, within $R_{\rm Lag}$, $\xi_{\rm c}$ is larger for less massive halos, but the mass trend reverses beyond $R_{\rm Lag}$:  i.e., less massive protohalos have steeper profiles.  This is in qualitative agreement with a generic prediction of peak theory \citep{Sheth1999}.  In the present context, this confirmation of the peak prediction is interesting, but it is only a means to an end.  

The cross power spectrum $P_{\rm c}(k)$ between the initial Lagrangian protohalo overdensity field $ \delta_{\rm halo} $, and that of the dark matter at the initial time, $\delta_{\rm m} $, is defined as 
\beq
 \langle \delta_{\rm halo }( \bm{k}_1 ) \delta_{\rm m}( \bm{k}_2 ) \rangle
 =  (2 \pi )^3  P_{\rm c }( k_1) \Ddel( \bm{k}_{1} +  \bm{k}_{2}  ) ,
 \label{eq:dhdm}
 \eeq
where  $\Ddel$ is the Dirac delta function.  Note that $P_{\rm c}(k)$ is the Fourier transform of $\xi_{\rm c}$, and that $\delta_{\rm halo}$ here should not be confused with $\delta_{\rm c}$, the overdensity within a protohalo patch, which played a major role in the previous section.  

Since the Lagrangian matter power spectrum is 
\beq
 \langle \delta_{\rm m}( \bm{k}_1 ) \delta_{\rm m}( \bm{k}_2 ) \rangle 
 =  (2 \pi )^3  P_{\rm m}( k_1) \Ddel( \bm{k}_{1} +  \bm{k}_{2} ) ,
\eeq
we define the Lagrangian cross bias parameter
\beq
\label{eq:bcL_k}
 b_{\rm c } (k, z)  \equiv 
 \frac{ D(z_{*}) }{ D(0) } \Big( \frac{P_{\rm c} (k, z_*;z)   }{ P_{\rm m}(k,z_*)  }-1 \Big)  ,
\eeq
where $D$ is the  linear growth factor.  Note that we extrapolate the Lagrangian bias parameter to $z=0$ using linear evolution of bias.  The presence of unity in Eq.~\ref{eq:bcL_k} is due to the finite initial redshift in simulation \cite[see, e.g.,][]{ChanScoccimarroSheth2012}. The argument $z$ denotes the redshift at which the Eulerian halos were identified. 

\begin{figure}
 \centering
 \includegraphics[width=0.9\linewidth]{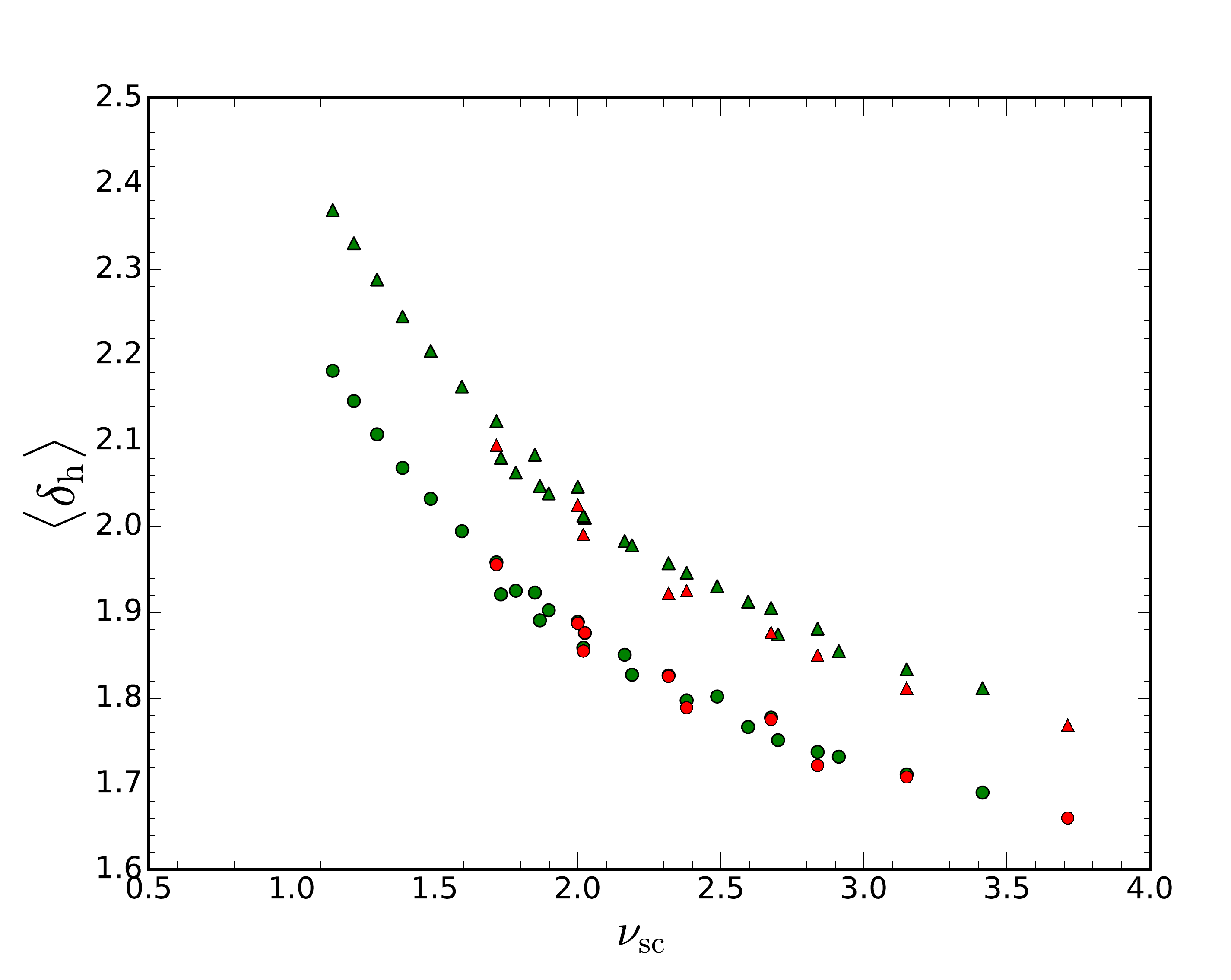}
 \caption{Direct estimates of the mean density enclosed within $R_{\rm Lag}$, shown as a function of the scaled mass variable $\nu_{\rm sc}\equiv\delta_{\rm sc}/\sqrt{s_0}$.  Green symbols show measurements based on Eq.~\ref{eq:Pk2dh} and red ones show the traditional estimator based on Eq.~\ref{eq:xic_WR}.  Data from the Carmen and Oriana simulation sets at $z=0.97$ and 0 are used.   The triangles (upper set) denote the estimates using $W_{\rm TH}$; the circles (lower set) used $W_{\rm Eff}$; in both cases, lower mass protohalos are more overdense.}
 \label{fig:dh_EffTH}
\end{figure}

\begin{figure}
 \centering
 \includegraphics[width=0.9\linewidth]{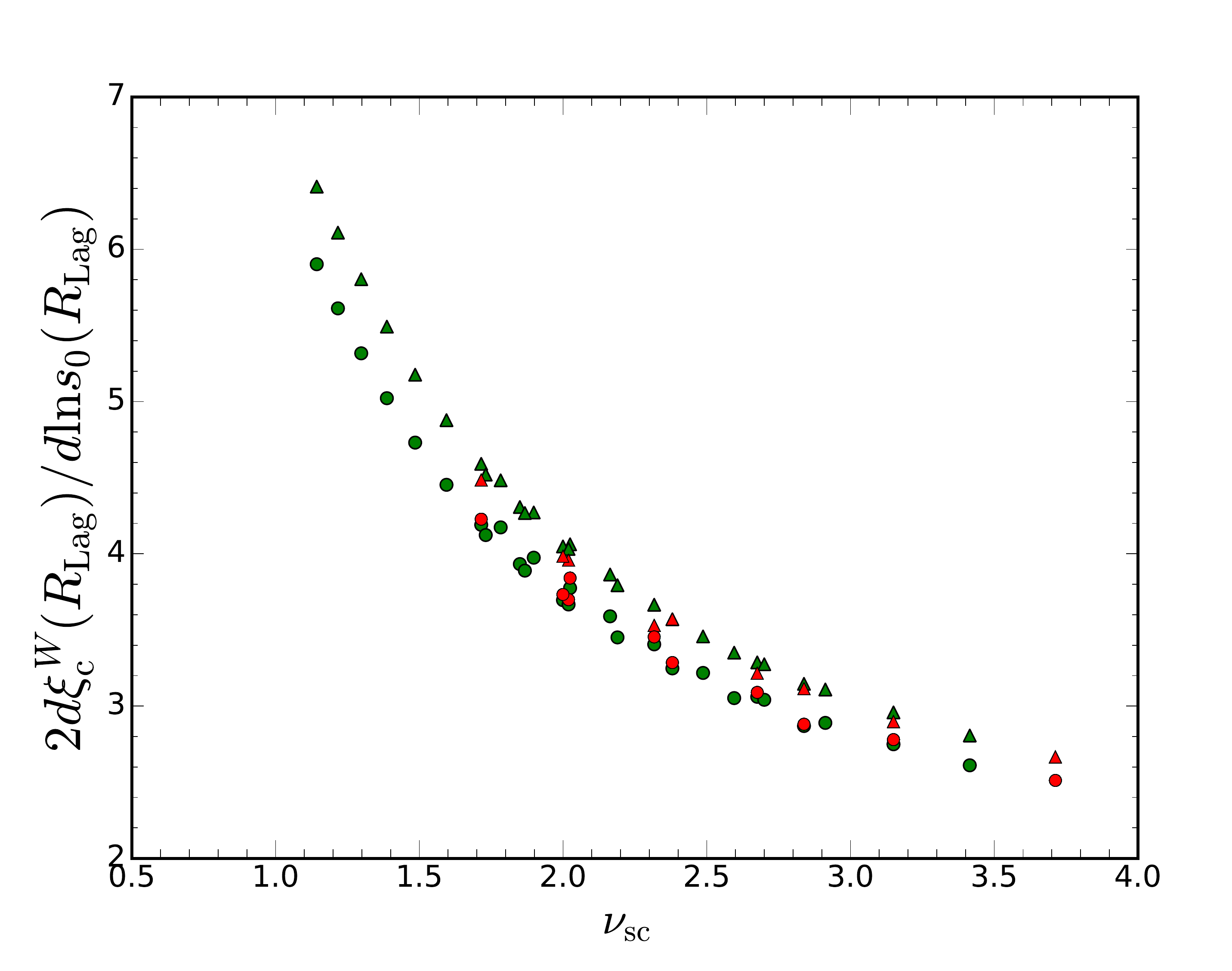}
 \caption{ Same as Fig.~\ref{fig:dh_EffTH}, but now for the slope variable $2 d \xi_{\rm c}^{\rm W}/d\ln s_0$.  Again, the configuration- and Fourier-based methods are in good agreement.}
 \label{fig:uh_EffTH}
\end{figure}

\subsection{Direct estimates of the mean enclosed overdensity and slope}
\label{sec:direct}
The consistency relations for the enclosed overdensity and its slope, both evaluated on the scale $R_{\rm Lag}$, state that
\begin{equation}
\label{eq:ConsistencyRelation_nu_u}    
 \left( \begin{array}{c}     
\nu_{\rm c}    \\
u_{\rm c}       \end{array}   \right) 
                         = \sqrt{s_0}   \left( \begin{array}{c}
                                               b_{1}^{(1)} + b_{1}^{(2)}   \\
                         \gamma_{\nu u} b_{1}^{(1)} + b_{1}^{(2)}/\gamma_{\nu u}  \end{array}   \right)
\end{equation}  
(compare Eq.~\ref{eq:ConsistencyRelation_2}).  We test these relations by measuring the quantities on the left- and right-hand sides.  


The enclosed overdensity is obtained by smoothing the density profile, so it is just 
\beq
\label{eq:xic_WR}
 \xi_{ \rm c }^{\rm W}(R)  =  \int dr\, 4 \pi r^2\, W(r;R)\, \xi_{\rm c}(r).
\eeq
Here $W$ is the inverse Fourier transform of the smoothing window defined earlier, so it has units of inverse volume.  E.g., for a tophat, $W = \Theta( R/r -1)/(4\pi R^3/3)$, and for $W_{\rm Eff}$, see \citet{ChanShethScoccimarro2015}. I.e., $\xi_{ \rm c }^{\rm W}(R)$ is the cross correlation $\xi_{\rm c}(r)$ shown in Fig.~\ref{fig:xic_rnorm}, smoothed by the effective window function. 

\begin{figure*}
\centering
\includegraphics[width=0.9\linewidth]{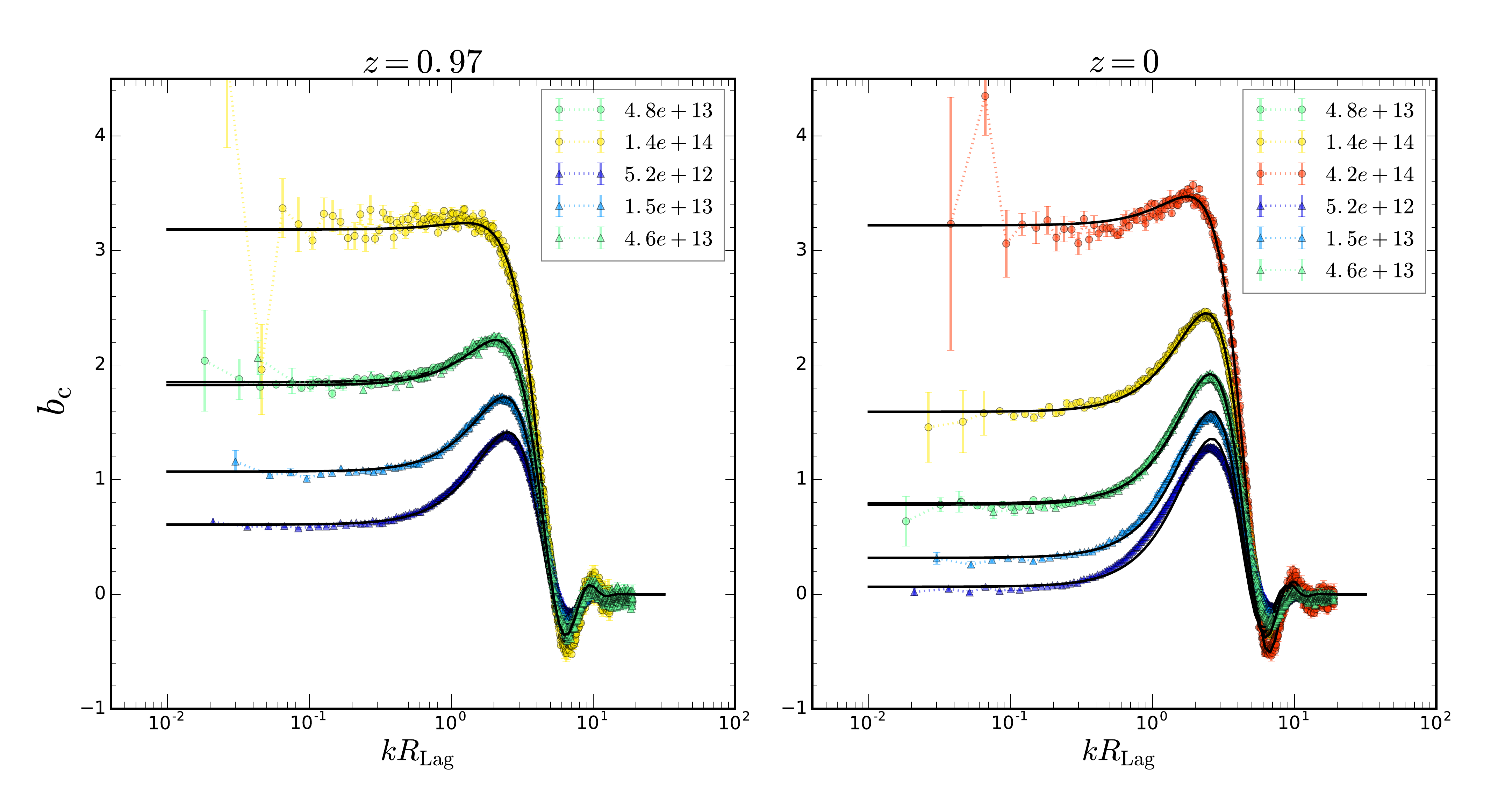}
\caption{ The bias parameter $b_{\rm c}$, obtained from the Fourier space cross-correlation between the dark matter and Lagrangian protohalos patches of halos identified at $z=0.97$ (left panel) and $z=0$ (right panel), in the Oriana (circles) and Carmen (triangles)  simulations.  Halos span a wide range of masses (legend is in units of $ \Msun $).  Smooth curves show the result of fitting Eq.~\ref{eq:bceff} to these measurements.}
\label{fig:bc_Lag_Dextrap}
\end{figure*}

Similarly, the slope variable is 
\beq
 \label{eq:Uh}
 \langle U_{\rm h}\rangle \equiv \langle d\delta_{\rm h}(R)/ds_0(R)\rangle = d\xi_{ \rm c }^{\rm W}(R)/ds_0(R),
\eeq
where $\delta_{\rm h}$ is the overdensity within a protohalo patch when smoothed on scale $R$ (it is not $\delta_{\rm halo}$ defined in Eq.~\ref{eq:dhdm}!).
For reasons which will become clear shortly, Fig.~\ref{fig:u_rnorm} shows a related measure of the slope: $2s_0\,\langle U_{\rm h} \rangle \equiv 2\,d\xi_{ \rm c }^{\rm W}(R)/d\ln s_0(R)$.  
As for $\xi_{\rm c}$, normalizing distances by $R_{\rm Lag}$ removes most of the mass dependence.  For all masses, we find that the slope has an obvious maximum on scales that are very close to $R_{\rm Lag}$, and this maximum is larger for the less massive halos.  Also note that there is an additional small dependence  on $z$. The maximum feature in Fig.~\ref{fig:u_rnorm} arises because there is an intrinsic window function in $\xi_{\rm c } $, and  $\xi_{\rm c }^{\rm W}  $ changes rapidly when the external window function has large overlap with the intrinsic one. This happens when  $R \sim R_{\rm Lag } $, and hence a peak results.

The quantities which are most directly related to the consistency relations are the mean enclosed overdensity and slope on scale $R_{\rm Lag}$.  While the previous expressions show how to estimate them from $\xi_{\rm c}$, they can also be written as sums over Fourier space quantities: 
\beq
 \label{eq:Pk2dh}
 \langle\delta_{\rm h}\rangle = \xi_{ \rm c }^{\rm W}(R_{\rm Lag}) 
 = \int \frac{dk\, k^2}{2\pi^2}\, P_{\rm c}(k)\,W(kR_{\rm Lag})
\eeq
and
\beq
 \label{eq:Pk2Uh}
 \langle U_{\rm h}\rangle 
 =  \frac{d\xi_{\rm c}^{\rm W}(R_{\rm Lag})}{ds_0(R_{\rm Lag})}
 = \int \frac{dk\, k^2}{2\pi^2}\, P_{\rm c}(k)\,\frac{dW(kR_{\rm Lag})}{ds_0(R_{\rm Lag}) }.
\eeq
Figs.~\ref{fig:dh_EffTH} and~\ref{fig:uh_EffTH} show that our direct estimates of $\langle\delta_{\rm h}\rangle$ and $\langle U_{\rm h}\rangle$ using $P_{\rm c}(k)$ are in excellent agreement with the more traditional estimators which are based on $\xi_{\rm c}(r)$.  Both estimators in Fig.~\ref{fig:dh_EffTH} show that the enclosed overdensity is larger for smaller masses (strictly speaking, for smaller $\nu_{\rm sc}$).  This is in agreement with previous direct measurements of the mean overdensity within protohalo patches \citep{ShethMoTormen2001,RobertsonKravtsovTinkerZentner2009,EliaLudlowPorciani2012,DespaliTormenSheth2013}.  

With these direct estimates of the overdensity and slope in hand, we are now ready to estimate the other quantities which appear in the consistency relations:  large-scale bias factors.  

\subsection{Lagrangian cross bias parameters in Fourier space}
\label{sec:crosspowerspectrum}

The symbols in Fig.~\ref{fig:bc_Lag_Dextrap} show the Lagrangian protohalo bias parameters $b_{\rm c}$ for halos identified at $z=0.97$ and 0 for a range of halo masses (as indicated) obtained using Eq.~\ref{eq:bcL_k} from the $P_{\rm c}$ measurements. Some of the mass dependence is removed because we plotted $b_{\rm c}$ against $kR_{\rm Lag}$ rather than $k$.  This plot is similar  to Fig.~5  in \cite{ChanShethScoccimarro2015}.  There we used it to highlight the fact that Eq.~\ref{eq:Weff} is crucial for estimating the Lagrangian bias parameters reliably, but we did not go into the details of the biasing model. Here we instead focus on the best fit bias parameters, and explore the consistency relations among them. 

\begin{figure*}
\centering
\includegraphics[width=0.9\linewidth]{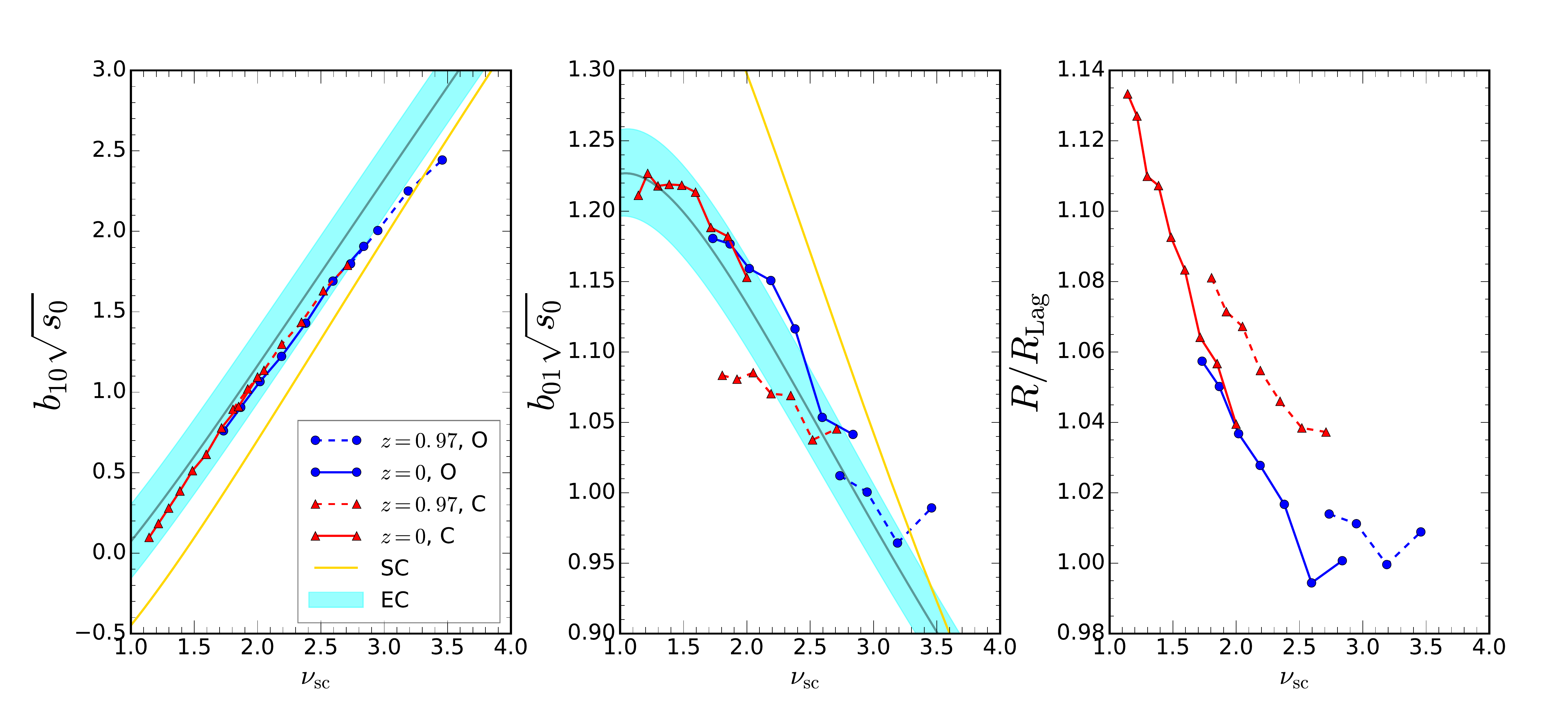}
\caption{Best-fit bias parameters $b_{10}$ and $b_{01}$, and window scale $R$ as a function of protohalo mass, shown as the scaled variable $\nu_{\rm sc} = \delta_{\rm sc}(z)/\sqrt{s_0(M)}$.  Symbols show measurements in the Oriana (blue circles) and Carmen (red triangles) simulations respectively, for halos identified at $z=0.97$ (dashed) and 0 (solid).   The panels for $b_{10}$ and $b_{01}$ also show predictions from the excursion set peak model  in which halos are assumed to form from a deterministic spherical collapse [($\delta_{\rm c} = \delta_{\rm sc}$) solid yellow curve]  or a stochastic ellipsoidal collapse in  \citep{ParanjapeShethDesjacques2013} (cyan band brackets $0.1 \le \beta \le 0.5$). }
\label{fig:bestfit_param_nu}
\end{figure*}

Our goal here is to model the measurements of $b_{\rm c}$ shown in Fig.~\ref{fig:bc_Lag_Dextrap}.  We start with the simple two-parameter model for the Lagrangian cross bias $b_{\rm c}$: 
\beq
\label{eq:bceff} 
 b_{\rm eff} (k ) = b_{10}\, W(kR) + b_{01} \, 2\frac{  d W(kR) }{d \ln s_0(R) } .
\eeq
As we mentioned, the two bias parameters $b_{10}$ and $b_{01}$ arise from the density threshold and slope constraints \citep[][except that our $b_{01}$ is what they call $b_{11}$]{MussoParanjapeSheth2012}.  We have changed notation from  $b_1^{(1)}$ in Sec.~\ref{sec:Theory} to $b_{10}$, etc., to highlight the fact that the biases in Sec.~\ref{sec:Theory} are un-averaged, while those measured in the simulations are averaged over the constraints as in Eq.~\ref{eq:Delta_averageC}. 
 
The smooth black curves in Fig.~\ref{fig:bc_Lag_Dextrap} show the result of treating $b_{10}$, $b_{01}$ and  $R$ as free parameters when fitting Eq.~\ref{eq:bceff} to the measurements.  In practice, to ensure that the low-$k$ part, which has large error bars relative to the high $k$-region, is properly fitted,  we first determine $b_{10}$ by fitting to the low $k$ constant part up to  $kR_{\rm Lag } <0.15$.   We can do this because $W\to 1$ and the scale-dependent term vanishes at $k\ll R_{\rm Lag}$ (c.f. Fig.~\ref{fig:deg_terms_compare}).   We then keep the best-fit  $b_{10}$  fixed and fit the remaining parameters $b_{01}$ and $R$. Evidently, this simple model for $b_{\rm eff }$ is able to provide a good fit over the entire range of $k$. 

Fig.~\ref{fig:bestfit_param_nu} shows the best-fit parameters, $b_{10}$, $b_{01}$, and $R$ as a function of halo mass, expressed in terms of $\nu_{\rm sc} = \delta_{\rm sc}(z)/\sqrt{s_0(M)}$ where large $M$ has large $\nu_{\rm sc}$).  Note that $s_0$ uses Eq.~\ref{eq:Weff} for the smoothing window, and the $z$ dependence of $\delta_{\rm sc}$ allows us to easily compare measurements for halos identified at different redshifts (in this case $z=0.97$ and 0).  The bias parameters from different $z$ coincide with each other rather well, in agreement with previous work \citep{ShethTormen}.  We find that, although the best fit $R$ is close to $R_{\rm Lag}$, the best-fit $R/R_{\rm Lag}$ decreases as $\nu_{\rm sc}$ increases.  For low $\nu_{\rm sc}$, especially $\nu_{\rm sc}\lesssim 1.5 $, the model does not work very well.  There are additional small differences between the results from Carmen and Oriana simulations.  

In principle, we should be able to use our Fourier space estimates of $P_c$ (essentially the solid curves in Fig.~\ref{fig:bc_Lag_Dextrap}) to predict the measured $\xi_{\rm c}$ of Fig.~\ref{fig:xic_rnorm}.  Since this is not the main focus of our study, we only show the result in Fig.~\ref{fig:bc_rnorm}.

In addition, it happens that our measurements of $b_{10}$ and $b_{01}$ are reasonably well described by the excursion set peak formulae of \cite{ParanjapeShethDesjacques2013} provided one assumes halos formed from an ellipsoidal collapse with some stochasticity.   In that model, a stochastic term proportional to $\sqrt{s_0} $ is added to the spherical collapse barrier so that the barrier increases as $\nu_{\rm sc} $ decreases.  The amount of stochasticity is controlled by the free parameter $\beta$.   The cyan band shows the range spanned by their models with $0.1\le \beta\le 0.5$.  The yellow curve shows $\beta=0$ (spherical collapse and no stochasticity).  While the general agreement is encouraging, constraining specific collapse models is not the main focus of our study \cite[for more discussion of models, see][]{CastorinaParanjapeHahnSheth2016}.  Rather, our goal is to test the accuracy of the consistency relation, Eq.~\ref{eq:ConsistencyRelation_nu_u}.

\subsection{Consistency relation for the smoothed enclosed overdensity}\label{sec:consist_rel_density}
\begin{figure}
\centering
\includegraphics[width=\linewidth]{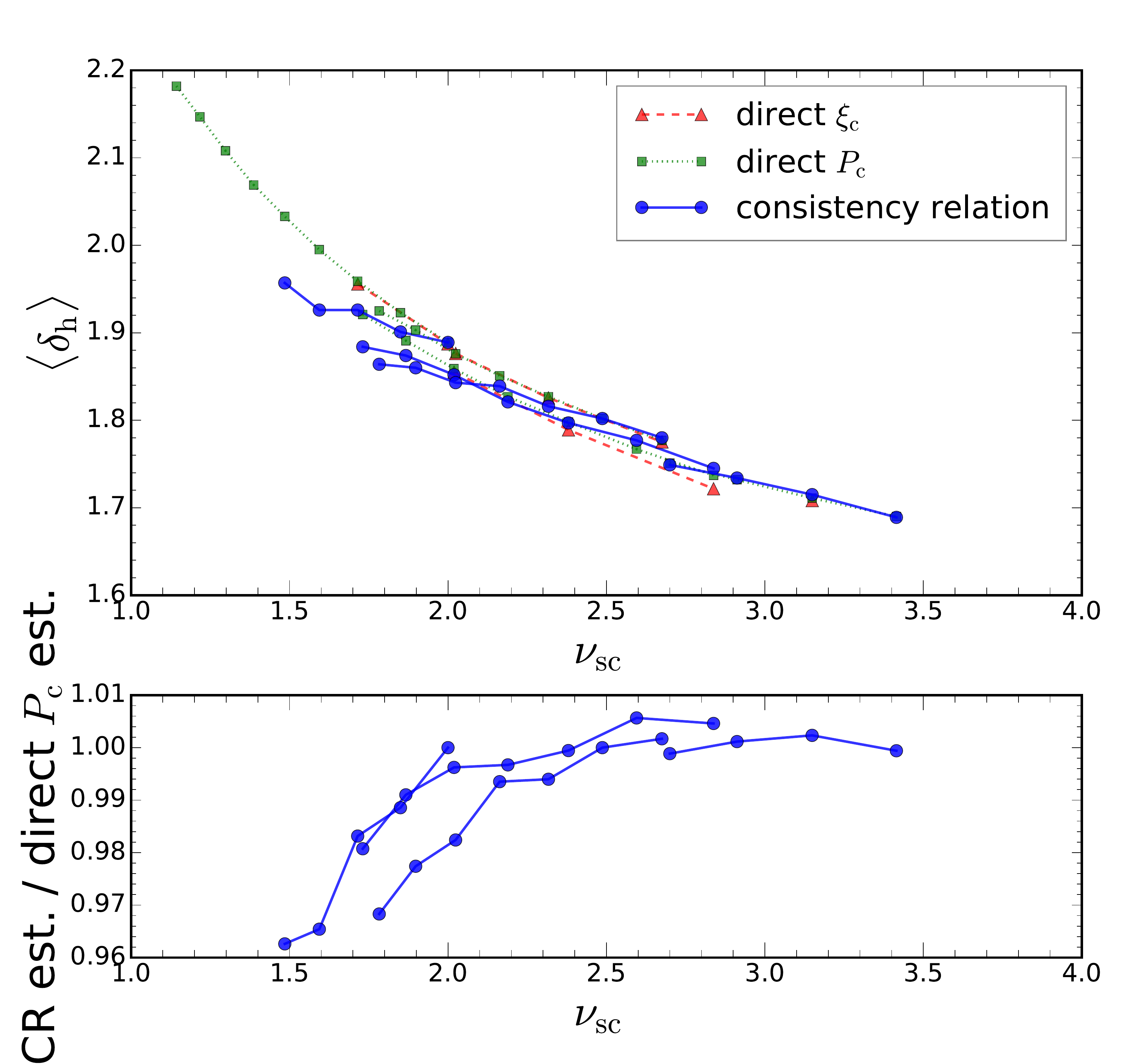}
\caption{Test of the first consistency relation (for protohalo overdensities).  Top panel compares direct measurements of  $\langle \delta_{\rm h} \rangle $, the enclosed overdensity in protohalo patches, with the consistency relation estimate (Eq.~\ref{eq:ConsistencyRelation_nu_u}).  Red triangles (dashed)  and green squares (dotted) show $\langle\delta_{\rm h}\rangle$ obtained  using real and Fourier-space methods (Eqs.~\ref{eq:xic_WR} and~\ref{eq:Pk2dh}) respectively.  Clearly, less massive halos formed from protohalo patches which were more overdense.  Blue circles show that the corresponding estimate from large scale bias $(b_{10} + b_{01})\, s_0$. Bottom panel shows the ratio between the consistency relation estimate and the direct estimate (from $P_{\rm c}$); they agree to better than 4 percent.}
\label{fig:bij_consistency_WeffPara_nu}
\end{figure}

The first consistency relation of Eq.~\ref{eq:ConsistencyRelation_nu_u} equates the sum of the large scale bias factors to $\langle\delta_{\rm h}\rangle/s_0$, where $\langle\delta_{\rm h}\rangle$ is the mean overdensity within the protohalo patch.  The red and green symbols in the upper panel of Fig.~\ref{fig:bij_consistency_WeffPara_nu} show our direct estimates of $\langle\delta_{\rm h}\rangle$ using Eq.~\ref{eq:xic_WR} and Eq.~\ref{eq:Pk2dh} with $W_{\rm Eff}$  (i.e., they are the lower set of symbols in Fig.~\ref{fig:dh_EffTH}).  The blue symbols and curves show our estimate of $\langle\delta_{\rm h}\rangle$ from the large scale bias factors:  $(b_{10}+b_{01})\,s_0(M)$. The lower panel shows that the consistency relation estimate of $ \langle \delta_{\rm h} \rangle $ is within 4\% of the direct estimate (we used the one shown by the green symbols).   This agreement is remarkable, given that the two estimates have {\em very} different systematics: that based on $b_{10}$ and $b_{01}$ is derived from two-point measurements on relatively large scales, whereas the traditional view of the direct measure is that it is more like a one-point measurement on substantially smaller scales.  We will have more to say about this agreement in Section~\ref{sec:revisit}.  For now, we simply note that the consistency relation has used just $n=2$ large scale bias factors to correctly predict the enclosed Lagrangian overdensity of the protohalos measured on much smaller scales.  

It is possible that the slight discrepancy which is apparent at lower $\nu_{\rm sc}$ is indicating that our model, which assumes there are only two important parameters, is overly simplistic.  We argued that, in principle, our analysis permits one to add as many parameters as desired to the bias prescription, so that the effective bias agrees with the simulation results well. That our simple model returns an estimate which is within a few percent of the direct measurement is a non-trivial self-consistency check that the simplest requirements (recall we have just two bias parameters) already capture most of the effects of bias.  The agreement shown in Fig.~\ref{fig:bij_consistency_WeffPara_nu} means that consistency relations have opened the door to using the scale dependence of bias to constrain halo formation physics.  Encouraged by this result, we now study the consistency relation associated with the second variable: the slope.

\subsection{Consistency relation for the slope}\label{sec:consistency_rel_firstcorssing}
We will now study the consistency relation for the slope $u$ rather than the enclosed overdensity $\delta_{\rm h} $.  A little algebra shows that the second of Eqs.~\ref{eq:ConsistencyRelation_nu_u} can be re-written as 
\beq
\label{eq:consistency_relation_u_t} 
 \gamma_{\nu u} \,  b_{10}  +  \frac{ b_{01}}{ \gamma_{\nu u} } 
   = \frac{ \langle u \rangle }{ \sqrt{ s_0 }  }
   = 2\gamma_{\nu u}\, \frac{d \xi_{ \rm c }^{\rm W}(R)}{ds_0(R)},
\eeq
where $u$ is the slope variable normalized by its rms value, defined in Eq.~\ref{eq:u_def}, and $\xi_{\rm c }^{\rm W}(R)$ is the enclosed overdensity defined in Eq.~\ref{eq:xic_WR} evaluated on scale $R_{\rm Lag}$ (also see Eq.~\ref{eq:Pk2dh}), so $d\xi_{ \rm c }^{\rm W}/ds_0$ is the original unnormalized slope (Eq.~\ref{eq:Pk2Uh}).  

\begin{figure}
\centering
\includegraphics[width=\linewidth]{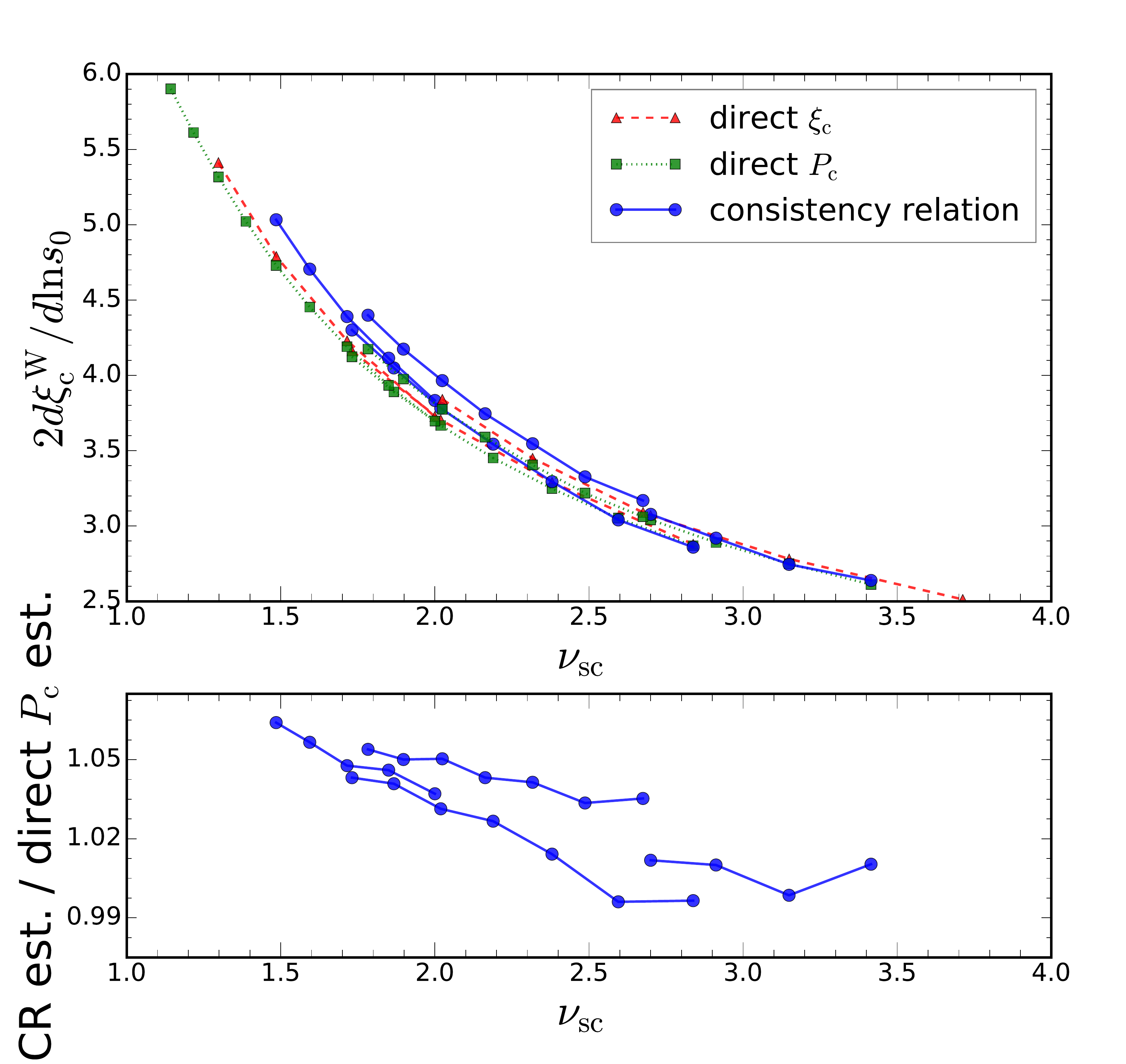}
\caption{  Test of the second consistency relation in Eq.~\ref{eq:ConsistencyRelation_nu_u}.  Colors and protohalo samples are the same as Fig.~\ref{fig:bij_consistency_WeffPara_nu}, but now show the slope of the smoothed, enclosed density protohalo profile rather than its magnitude as a function of scaled protohalo mass. In the upper panel, blue circles show the estimate using $(b_{10}+b_{01}/\gamma_{\nu u}^2)\,s_0$, and red triangles and green squares show real and Fourier-based direct measurements of the mean slope. The ratio between the consistency relation estimate and the direct estimate using  $P_{\rm c} $  in the lower panel shows that the level of agreement between these estimates is similar to that for the enclosed density.   
}
\label{fig:bij_consistency2_crossing}
\end{figure}

The first consistency relation states that $s_0$ times the sum of the bias factors is an estimator of $\delta_{\rm h}$.  For similar reasons, it is interesting to view the second consistency relation as equating $s_0\,(b_{10} + b_{01}/\gamma_{\nu u}^2)$ to $2\,d\xi_{\rm c}^W/d\ln s_0$ on scale $R_{\rm Lag}$.  We take this latter quantity from the direct measurements shown in Fig.~\ref{fig:u_rnorm}; these are shown as the red curves in the upper panel of Fig.~\ref{fig:bij_consistency2_crossing}.  The green symbols and curves show the Fourier-based estimator of Eq.~\ref{eq:Pk2Uh}.  These clearly show steeper slopes for lower mass protohalos, which we asserted earlier was a generic prediction of peak theory.  The blue curves show $s_0\,(b_{10} + b_{01}/\gamma_{\nu u}^2)$.  The large scale bias factors have correctly estimated the trend for lower mass protohalos to have steeper slopes on the scale $R_{\rm Lag}$. The ratio between the consistency relation estimate and the direct measurement using $P_{\rm c}$ is plotted in the lower panel of  Fig.~\ref{fig:bij_consistency2_crossing}.  The agreement between these estimates is roughly  as good as it was for the enclosed overdensity (compare Fig.~\ref{fig:bij_consistency_WeffPara_nu}). 

Figs.~\ref{fig:bij_consistency_WeffPara_nu} and \ref{fig:bij_consistency2_crossing} are the main results of this paper.  They show that the scale dependence of large-scale bias provides reliable estimates of both the density and its slope at the (smaller) Lagrangian radius -- quantities which are expected to encode the small scale physics of halo formation.  Note in particular that these estimates do not require a priori knowledge of the physics of collapse (spherical or not, deterministic or not), or the nature of the halo population (assembly biased or not).  

\subsection{Consistency relations revisited}
\label{sec:revisit}
Eqs.~\ref{eq:Pk2dh} and~\ref{eq:Pk2Uh} provide a simple way to see why the consistency relations work.  Start with $P_{\rm c}(k) = b_{\rm c}(k)\,P_{\rm m} (k)$ and assume that Eq.~\ref{eq:bceff} for $b_{\rm c}$ is not just accurate, it is exact.  Then 
\begin{align}
 \label{eq:d_exact}
 \langle\delta_{\rm h}\rangle &= \int \frac{dk\, k^2}{2\pi^2}\, P_{\rm m}(k)\,
        \left[ b_{10}\,W^2 + b_{01}\,2W\frac{dW}{d\ln s_0} \right] \nonumber\\
 &= \int \frac{dk\, k^2}{2\pi^2}\,P_{\rm m} (k)\left[b_{10}\,W^2 + b_{01}\,\frac{dW^2}{d\ln s_0}\right] \nonumber\\
 &= b_{10} s_0 + b_{01} \, \frac{ds_0}{d\ln s_0} = s_0\,(b_{10} + b_{01}),
\end{align}
which is the first of the consistency relations in Eq.~\ref{eq:ConsistencyRelation_nu_u}.  It is a simple matter to verify that if $b_{\rm c}=b_{\rm eff}^{(3)}$ of Eq.~\ref{eq:bceff_3} then the expression above for $\langle\delta_{\rm h}\rangle$ becomes the first of the consistency relations in Eq.~\ref{eq:ConsistencyRelation_3}. This shows that, if our model for $b_{\rm c}$ is correct, then the consistency relation is a tautology.  

A similar analysis of the slope, i.e., inserting Eq.~\ref{eq:bceff} in Eq.~\ref{eq:Pk2Uh}, yields 
\begin{align}
 \label{eq:u_exact}
 2s_0\,\langle U_{\rm h}\rangle &= \int \frac{dk\, k^2}{2\pi^2}\,P_{\rm m} (k)\,2\frac{dW}{d\ln s_0}
        \left[ b_{10}\, W + b_{01}\,2\frac{dW}{d\ln s_0} \right] \nonumber\\
 &= s_0\, b_{10} + b_{01} 4 s_0^2 s_u = s_0\,(b_{10} + b_{01}/\gamma^2_{\nu u})
\end{align}
which is the second relation in Eq.~\ref{eq:ConsistencyRelation_nu_u}.  And if $b_{\rm c}=b_{\rm eff}^{(3)}$, then the expression above for $2s_0\,\langle U_{\rm h}\rangle$ becomes the third of the consistency relations in Eq.~\ref{eq:ConsistencyRelation_3}.  

Now suppose we have only reliable knowledge about the window function for $k\le k^\dag $.  Then it is useful to define $W^{\dag}$, which is the same as the true window function $W$ up to $k^\dag$, beyond which it is set to zero. Then although $P_{\rm c}$ is still given by the true window function $W$, the enclosed overdensity estimated using $W^{\dag}$ is 
\begin{align}
 \langle\delta_{\rm h}^{\dag} \rangle &= \int \frac{dk\, k^2}{2\pi^2}\,  P_{\rm m}(k) \,
       W^{\dag}   \left[ b_{10}\,W + b_{01}\,2\frac{dW}{d\ln s_0} \right]   \nonumber\\
 &= b_{10} s_0^{\dag} + b_{01}  \int \frac{dk\, k^2}{2\pi^2}\,  P_{\rm m}(k)  \,  \frac{ d (W^{\dag} )^2  }{  d \ln s_0^\dag } \frac{ d \ln s_0^\dag   }{ d \ln s_0 }   \nn \\
 & = s_0^\dag \Big( b_{10} + b_{01}^\dag \Big) ,  
 \label{dWdagger}
\end{align}  
where $ s_0^\dag $ is computed using  $W^{\dag}$ and $b_{01}^\dag \equiv b_{01}\,(d \ln s_0^\dag/d \ln s_0)$.  Similarly the consistency relation for the slope becomes 
\begin{align}
 2s_0^\dag \,\langle U_{\rm h}^\dag \rangle &= \int \frac{dk\, k^2}{2\pi^2} \,P_{\rm m} (k)  \,2\frac{dW^\dag}{d\ln s_0^\dag}
        \left[ b_{10}\, W + b_{01}\,2\frac{dW}{d\ln s_0} \right]   \nonumber\\
       & = s_0^\dag \left( b_{10}  + \frac{ b_{01}^\dag }{ ( \gamma_{\nu u }^\dag )^2  } \right),  
\end{align}
where $ \gamma_{\nu u }^\dag $ is also computed using  $W^\dag$.  This shows that both consistency relations look like the original ones, with rescaled $s_0$ and $\gamma_{\nu u}$, and $b_{01}$ values.  

To see what this means, consider what happens when we estimate $b_{\rm c}$ from the measured $P_{\rm c}$ by fitting to $k\le k^\dag$.  Suppose that the true scale dependent bias piece is $b_{01}\,dW/d\ln s_0$.  If we are unsure of the high-$k$ part of $W$, then we cannot compute $s_0$, and hence $dW/d\ln s_0$ when fitting to determine $b_{01}$.  This means that when we fit, we are forced to fit for some $b_{01}^\dag$ using $dW^\dag/d\ln s_0^\dag$.  Getting a good fit means
    $b_{01}^\dag\, dW^\dag/d\ln s_0^\dag = b_{01}\, dW/d\ln s_0$
and we are assuming $W^\dag = W$ for $k \le k^\dag$, the result of fitting $b_{\rm c}$ well will be that
    $b_{01}^\dag \equiv b_{01}\,(d \ln s_0^\dag/d \ln s_0)$. 
Blindly inserting this in the consistency relation yields Eq.~\ref{dWdagger}.  

Thus, provided we have fit $b_{\rm c}$ well, and provided we use $W^\dag$ everywhere, the consistency relation will be satisfied.  However, $\langle \delta^\dag_{\rm h} \rangle$ will not be the same as $\langle\delta_{\rm h}\rangle$ obtained with the true smoothing window.  In addition, as we change $k^\dag$, $d\ln s_0^\dag/d\ln s_0$ will change.  As a result, $b_{01}^\dag$ will depend on the value of $k^\dag$, until $d\ln s_0^\dag/d\ln s_0$ is close enough to 1 that the difference doesn't matter.  I.e., the difference between the true $W$ and $W^\dag$ does not matter for $k$ larger than this final $k^\dag$.  If $k^\dag R_{\rm Lag} < 1$, then we will truly have used the large scale information (on scale $R^\dag \sim 1/k^\dag$) to constrain the density on scale $R_{\rm Lag}$.  
Note that the requirement is not really $|d\ln s_0^\dag/d\ln s_0| - 1 \ll 1$.  Rather all we really need is $b_{10} + b_{01}^\dag$ to be close enough to $b_{10} + b_{01}$.  For $b_{10} \gg b_{01}$, it may be that one can live with fairly small $k^\dag$ and still be OK.  The requirement for the slope is slightly more stringent, since then we also need $\gamma_{\nu u}^\dag$ to have converged to $\gamma_{\nu u}$.  

The analysis above makes the point that the accuracy with which the consistency relation estimates the smoothed overdensity is directly related to how well our model for $b_{\rm c}$ actually fits the true $b_c$; this depends in part on how well the assumed form for $W$ approximates the true $W$.  Although one might have thought that many parameters were required to achieve accurate results, Figs.~\ref{fig:bc_Lag_Dextrap}--\ref{fig:bij_consistency2_crossing} indicate that the simple two-parameter model, motivated by the excursion set approach, is good enough for reconstructing the enclosed overdensity and its slope.

It turns out to be straightforward to extend this analysis to higher order bias.  Start with Eq.B2 in \cite{MussoParanjapeSheth2012}, which states that, in a Gaussian random field, 
\begin{align}
  & \xi_n^{\rm W}(R_0) \equiv \frac{\big[s_0(R_0)\big]^{n/2}}{N_{\rm h}}  \sum_{i=1}^{N_{\rm h}} H_n\Bigl(\frac{ \delta^{\rm W}_i(R_0)}{\sqrt{ s_0(R_0) }}\Bigr) \\
 &= \int \prod_{i=1}^n\frac{d\bmath{k}_i}{(2\pi)^3} 
                   \,P_{\rm m} (k_i)\, W(k_iR_0)  W(k_i R_{\rm Lag})\, c_n(\bmath{k}_1,\ldots ,\bmath{k}_n),                   
 \label{eq:MPS}
\end{align}
where $\delta^{\rm W}_i(R_0)$ is the matter overdensity at the position of the $i$th halo, smoothed on scale $R_0$ with window $W$, and $H_n$ is the probabilist's Hermite polynomial, and the $c_n$ are $n$th order bias coefficients.  (Strictly speaking, they wrote the sum over halos which appears in the first equality above as $\langle(1+\delta_{\rm h})H_n\rangle$.  Since $\delta_{\rm h}$ means something else here, we have simply written the sum explicitly.  In addition, our expression corrects a few other typos in theirs.)  Our notation highlights the fact that $\xi_n^{\rm W}$ is a cross correlation between protohalo positions and Hermite-polynomial weightings of the smoothed density fluctuation field.  In a model where only $\nu$ and $x$ matter, 
\begin{align}
 c_n(\bmath{k}_1,\ldots ,\bmath{k}_n) 
 & = b_{n  , 0} + b_{n-1  , 1}\sum_{i=1}^n \frac{s_0}{s_1}\,k_i^2  \nn \\
 &  + b_{n-2  , 2}\,\sum_{i<j}\,\frac{s_0}{s_1}k_i^2 \frac{s_0}{s_1}k_j^2 + \cdots .
   \label{eq:cnk}
\end{align}

When $R_0 = R_{\rm Lag}$ then it is easy to see that the result of inserting Eq.~\ref{eq:cnk} in Eq.~\ref{eq:MPS} is 
\beq
 \frac{\xi_n^{\rm W}(R_{\rm Lag})}{ \big[ s_0(R_{\rm Lag}) \big]^n } = \sum_{j=0}^n {n\choose j}\, b_{j , n-j} ;
\eeq
this is the consistency relation statement that a binomial coefficient weighting of the scale-dependent bias factors yields the averaged Hermite-polynomial of the smoothed-overdensity centered on the protohalos.  If we replace $x$ with $u$, then the $(s_0/s_1)\,k_i^2$ factors become $2\,d\ln W_i/d\ln s_0$ but our final expression for $\xi_n^{\rm W}(R_{\rm Lag})$ is unchanged.   In this case, $\xi_1^{\rm W}$ is just $\xi_{\rm c}^{\rm W}$ of the previous subsections.  It is easy to check that the relation for the slope also works out easily.  And finally, if $m$ parameters matter, then the binomial weighting becomes a multinomial \citep{CastorinaSheth2017}.

\section{ Conclusions} 
\label{sec:conclusion}

In the initial Lagrangian space, protohalo patches which are destined to evolve into halos at a later time must satisfy a number of constraints.  Typically, these constraints involve the value of the smoothed density field as well as its derivatives with respect to position and scale.  Each of these constraints leaves its mark on the spatial distribution of the protohalos:  associated with each constraint is a hierarchy of Lagrangian bias parameters (linear, second order, etc.).  We used a simple matrix algebra analysis to show that if there are $n$ constraints, there will be $n$ linear bias parameters which satisfy $n$ self-consistency relations (Section~\ref{sec:general}).  To date, only one of these $n$ relations has been highlighted.  Section~\ref{sec:revisit} provided a rather different Fourier-space analysis of these relations.  Both approaches show that these relations encode information about the constraints:  We demonstrated this using measurements of the properties and clustering of protohalo patches in the initial conditions of numerical simulations. 

Our consistency relations show that measurements of the scale-dependence of bias can be combined to estimate the critical overdensity required for halo formation.  This estimate shows that the critical overdensity, smoothed over the protohalo patch scale, should be larger for less massive halos, and is in excellent agreement with direct measurements of the overdensity within the protohalo patches (Fig.~\ref{fig:bij_consistency_WeffPara_nu}).  We used two estimators for the direct measurement: one is a closer to the traditional estimator, and is based on averaging in configuration space (Eq.~\ref{eq:xic_WR}).  The other is a new estimator built from Fourier space measurements (Eq.~\ref{eq:Pk2dh}).  The two are in excellent agreement (Fig.~\ref{fig:dh_EffTH}).  Our Fourier-based estimator provides a particularly transparent way to see why the consistency relations work so well (Eq.~\ref{eq:d_exact}).  

The scale-dependence of bias can also be used to estimate the slope of the overdensity around protohalos, evaluated on the scale of the patch.  This predicted slope is steeper around less massive protohalos, and is again in excellent agreement with direct measurements of the slope (Fig.~\ref{fig:bij_consistency2_crossing}).  As for the mean enclosed density, our direct measurements of the mean slope were also based on configuration- (Eq.~\ref{eq:Uh}) and Fourier-based (Eq.~\ref{eq:Pk2Uh}) methods, which agreed well (Fig.~\ref{fig:uh_EffTH}).  And the Fourier-based estimator again provides a simple route to the consistency relations (Eq.~\ref{eq:u_exact}).

Our formalism suggests that, although the values of the large scale bias factors may depend on the shape of the smoothing window, the combination which matters for the consistency relation does not.  In this respect, detailed a priori knowledge of the shape and scale of the smoothing is not necessary for our methodology to work.  Explicit measurements of this dependence confirm this expectation (Figs.~\ref{fig:compareWs} and~\ref{fig:slopeWs}).  

Thus, our analysis has shown that the large-scale bias parameters, which are usually regarded as nuisance parameters in cosmological analyses, actually encode useful information about the small-scale physics of halo formation.  The consistency relations we highlighted allow one to decode this information.  Although we illustrated our approach using a simple model in which only two (sensibly chosen!) parameters matter, we showed that our analysis carries over, essentially unchanged, when there are more bias parameters.  This is important because, as the required precision on our understanding of halo formation increases, we expect to encounter more and more constraints, and hence more bias parameters.  Our analysis showed how to proceed when some of these parameters are nearly degenerate with others.  Moreover, even in the two parameter case, our analysis is immune to assembly bias, in the sense that the same consistency relations can be used for any subset of the halo population, without any a priori information about the nature of the subset.  Indeed, instead of breaking down in the presence of assembly bias, the consistency relations allow us to learn about the physics of assembly bias.  


Our analysis leads to three interesting directions for further study.  The first is to extend our linear algebra method, which is particularly simple for understanding the relations between linear bias parameters, to determine the consistency relations associated with higher order bias terms.  (Our Fourier space approach was generalized to higher order bias in Section~\ref{sec:revisit}.) The second is to study how the consistency relations for the Lagrangian bias of protohalo patches which we have studied here are modified when one considers the Eulerian bias of evolved halos.  Finally, our results clearly have implications for studies which exploit the relation between the scale independent linear bias parameter $b_{10}$  and halo mass, and which seek to relate the amplitude of $b_{10}$ to higher order bias parameters ($b_{20}$, say, see e.g.~\cite{Hoffmann:2016omy}).  Clarifying these issues is the subject of work in progress.

\section*{Acknowledgements} 
We thank the participants of the workshop ``Statistics of Extrema in Large Scale-Structure'' at the Lorentz center for helpful feedback in March 2016, and V.~Desjacques for his hospitality in Geneva and the ICTP for its hospitality in Trieste during the summers of 2015 and 2016 where some of this work was completed.  We thank the LasDamas project\footnote{\url{http://lss.phy.vanderbilt.edu/lasdamas}} for the simulation outputs used in this work. The simulations were run using a Teragrid allocation; some RPI and NYU computing resources were also used. KCC acknowledges the support from the Swiss National Science Foundation and the Spanish Ministerio de Economia y Competitividad grant ESP2013-48274-C3-1-P.
 Finally, we thank the referee for a helpful report. 

\bibliographystyle{mnras}
\bibliography{references}

\appendix

\section{Cross correlations in configuration space}\label{sec:crosscorrelationfunction}
The main text uses bias parameters estimated from protohalo-mass cross correlations in Fourier space.   
An interesting by-product of our analysis is an understanding of the same cross correlations in configuration space.  We first use a simple analytically tractable case to illustrate how the choice of smoothing filter is expected to impact our analysis.  We then show configuration space measurements from our simulations, and illustrate how well they are reproduced by our Fourier space analysis.  This comparison has no free parameters, so any disagreement is a consequence of systematics.  Before illustrating how the choice of smoothing window affects our use of consistency relations in practice, we use our cross-correlation measurements to illustrate a point which was recently made by \cite{CastorinaParanjapeSheth2016}:  that suitably normalized cross-correlations with the slope of the large scale fluctuation field should yield the {\em same} large scale bias as cross-correlations with the density itself.  Finally, we show that our use of consistency relations does not require  detailed knowledge of the shape of the smoothing window.

\subsection{Dependence on smoothing window:  Theory}
Suppose that the power spectrum is a power law 
\beq
 \label{eq:Pk_inv2}
 P_{\rm m} (k) = Ak^{-2}. 
\eeq
See \cite{MussoSheth2014} for more discussion of why this is an interesting choice in the cosmological context.  We set $A=62.6 \MpcOh$ by matching the $z=0$ $\Lambda$CDM $P(k)$ at $k=1\hOMpc$.  The dark matter correlation function is  
\beq
 \label{pk2xir}
 \xi_{\rm m}(r)  = \int \frac{ d k }{(2 \pi)^3} 4 \pi k^2 \, P_{\rm m} (k)\,j_0 ( k r )
                 = \frac{A}{ 4 \pi r }. 
\eeq
In the upcrossing approximation of the excursion set approach, the protohalo-mass cross power spectrum is 
\beq
 \label{eq:Phm}
 P_{\rm c} (k) = \Big[ b_{10} W(kR) + 2 b_{01} \frac{dW(kR)}{d \ln s_0(R)} \Big]   P_{\rm m} (k),
\eeq
where $R$ is the smoothing scale of the window associated with defining protohalo patches, and $s_0(R)$ is given by Eq.~\ref{eq:sj}. 

For the top-hat window ($W = W_{\rm TH}$), the second term can be simplified by noting that 
\beq
 \frac{d W_{\rm TH}}{d \ln s_0 } =  \frac{d \ln R }{ d \ln s_0 }  \frac{ d W_{\rm TH} }{d \ln R  }  =  3( W_{\rm TH} - j_0 ),
\eeq
because $d\ln R / d\ln s_0 = -1 $ for Eq.~\ref{eq:Pk_inv2}.  Setting $W=W_{\rm TH}$ in Eq.~\ref{eq:Phm} and inverse Fourier transforming yields 
\beq
 \xi_{\rm c} =  \frac{A}{ 2 \pi^2 }   \int d k j_0( k r ) \Big[ ( b_{10} + 6b_{01} ) W_{\rm TH} ( k R )   -  6 b_{01}j_0( k R)   \Big] .  
\eeq
If we define $\rho \equiv  r / R $, then  
\beq
\label{eq:xi2hh}
\xi_{\rm c}( R \rho)   =\begin{cases} \xi_{\rm m}(R)  \Big[  \frac{ 3 -\rho^2 }{ 2 } b_{10} + 3 (1 - \rho^2 )b_{01}   \Big]     & \mbox{if }  0 \le \rho \le 1  \\
                       b_{10} \xi_{\rm m}( R \rho)              & \mbox{if }  \rho  > 1
              \end{cases} . 
\eeq
Remarkably, for $\rho\ge 1$, the ratio $\xi_{\rm c}/\xi_{\rm m}$ is constant and equal to $b_{10}$.  I.e., the configuration space bias is independent of scale for all $r\ge R$; it becomes scale dependent only when $r < R$.  

\begin{figure}
\centering
\includegraphics[width=0.9\linewidth]{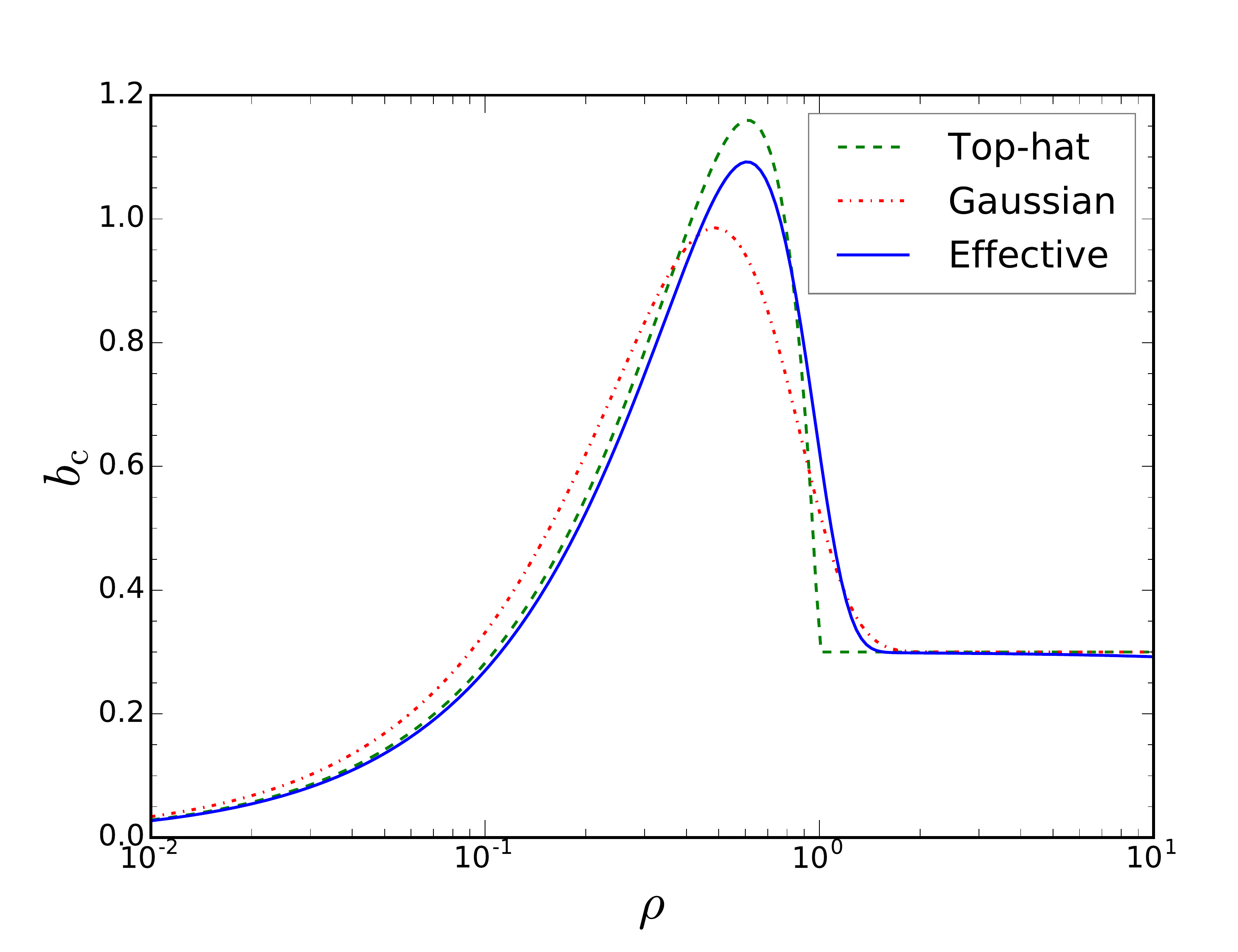}
\caption{ Cross-bias parameter in configuration space obtained with top-hat (dashed, green), Gaussian (dotted-dashed, red), and effective (solid, blue) smoothing windows respectively, when $P(k)\propto k^{-2}$.  }
\label{fig:bc_TH_G_Eff}
\end{figure}

\begin{figure}
\centering
\includegraphics[width=0.9\linewidth]{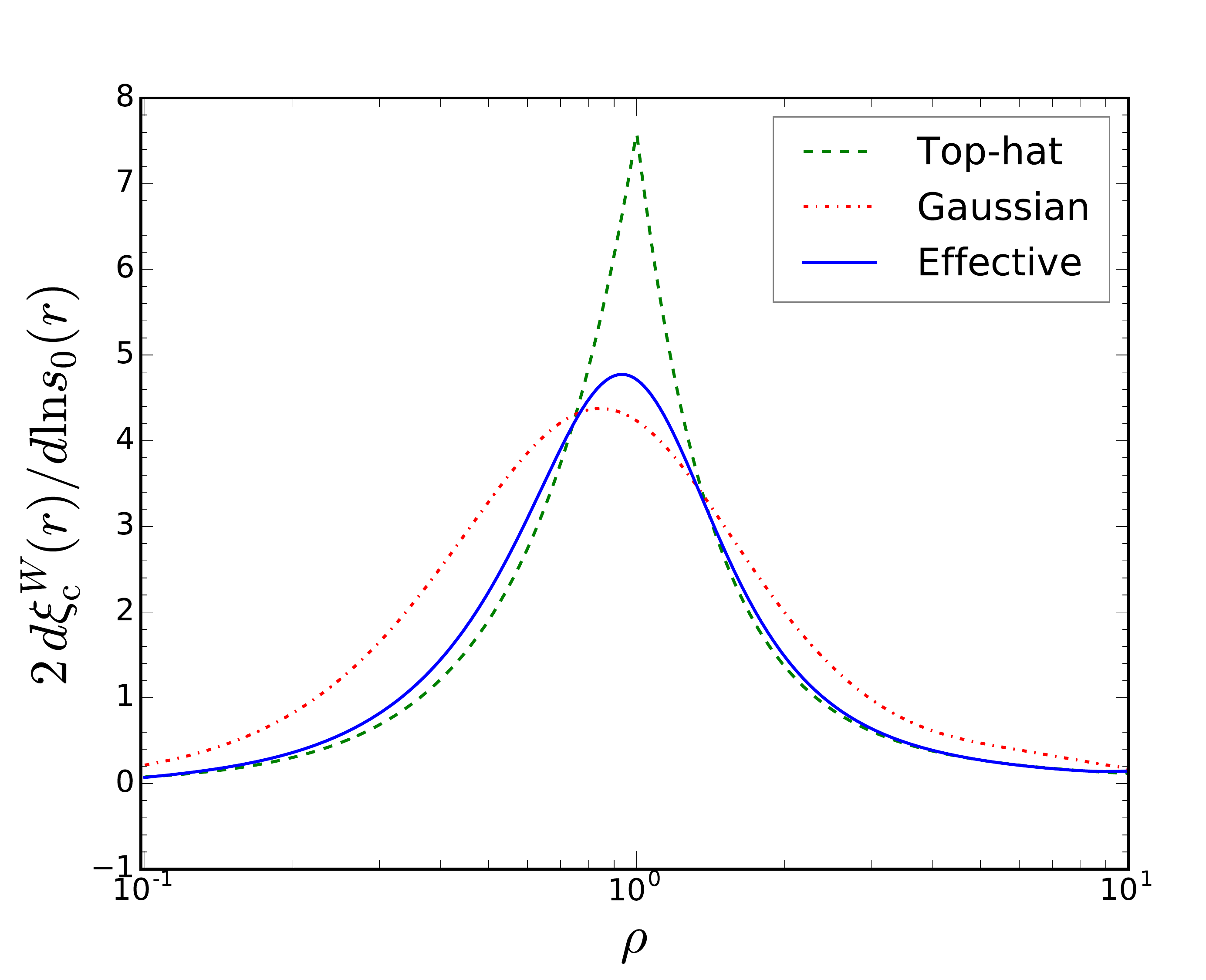}
\caption{ $  2 \, d \xi_{\rm c}^W(r) / d \ln s_0(r) $  obtained with top-hat (dashed, green), Gaussian (dotted-dashed, red), and effective (solid, blue) smoothing windows respectively, when $P(k)\propto k^{-2}$.  }
\label{fig:dxic_dlns0_2_TH_G_Eff}
\end{figure}

\begin{figure*}
\centering
\includegraphics[width=0.9\linewidth]{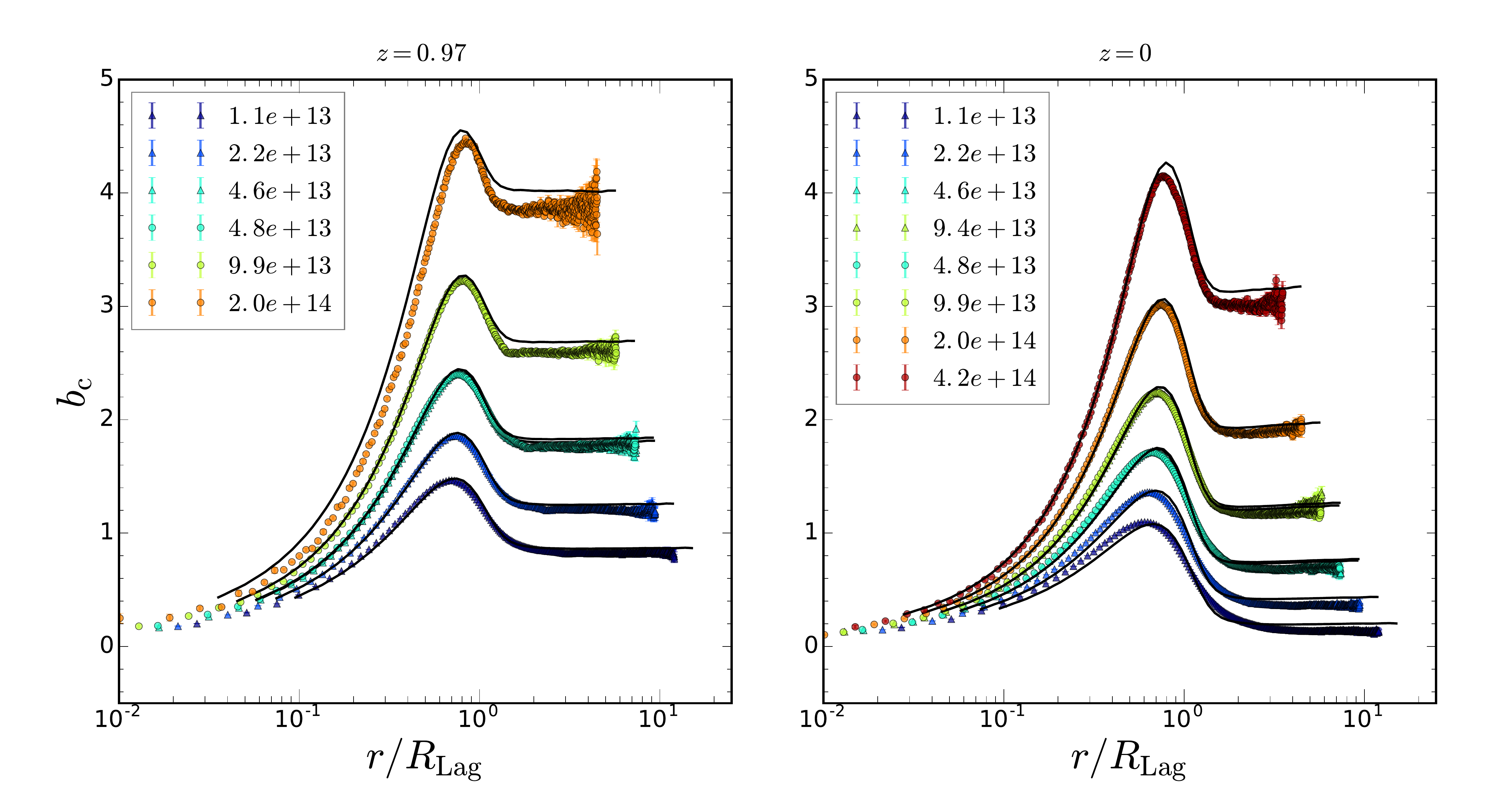}
\caption{ Bias factor $ b_{\rm c}$ obtained from inserting the measurements shown in Fig.~\ref{fig:xic_rnorm} into Eq.~\ref{eq:bcL_r}.  Solid black lines are not fits:  they show the predicted scale dependent bias associated with using the Fourier space estimates of $b_{10}, b_{01}$ and $R$ shown in Fig.~\ref{fig:bestfit_param_nu} in Eq.~\ref{eq:bpk2bxir}, and using the resulting $\xi_{\rm c}^{\rm thy}$ in place of the measurements in Eq.~\ref{eq:bcL_r}.  }
\label{fig:bc_rnorm}
\end{figure*}

In contrast, for Gaussian smoothing,
\beq
\label{eq:Phm_G}
P_{\rm c} (k) = \Big[ b_{10} W_{\rm G}(kR_{\rm G} ) + 2 b_{01} \frac{d  W_{\rm G}(kR_{\rm G})  }{d \ln s_0(R_{\rm G}) } \Big]  P_{\rm m} (k),   
\eeq
where $R_{\rm G}$ is the smoothing scale for the Gaussian window.  We set $R_{\rm G} = R/\sqrt{5}$ to 
match $W_{\rm G}$ and $W_{\rm TH}$ to lowest order in $k$.  Since 
\beq
 \frac{d  W_{\rm G} (kR_{\rm G})  }{d \ln s_0 (R_{\rm G})} = (k R_{\rm G}  )^2 W_{\rm G} (kR_{\rm G}),
\eeq
we have 
\beq
\label{eq:xihm_form}
\xi_{\rm c} (r) 
 =  b_{10}\, \xi_{\rm m}(r)\, \mathrm{ erf}\Big(  \frac{\rho_{\rm G} }{\sqrt{2}}  \Big) + 4\, b_{01} \xi_{\rm m} (R_{\rm G}) \frac{e^{-\rho^2_{\rm G}/2}}{\sqrt{2\pi}}  , 
\eeq
where $\rho_{\rm G} \equiv r / R_{\rm G}  $.  Clearly, for this filter, $\xi_{\rm c}/\xi_{\rm m}$ is scale dependent on all scales.  This is also true for $W_{\rm Eff}$ (Eq.~\ref{eq:Weff}) which we used extensively in the main text (and which we treat numerically).  

\begin{figure}
\centering
\includegraphics[width=0.9\linewidth]{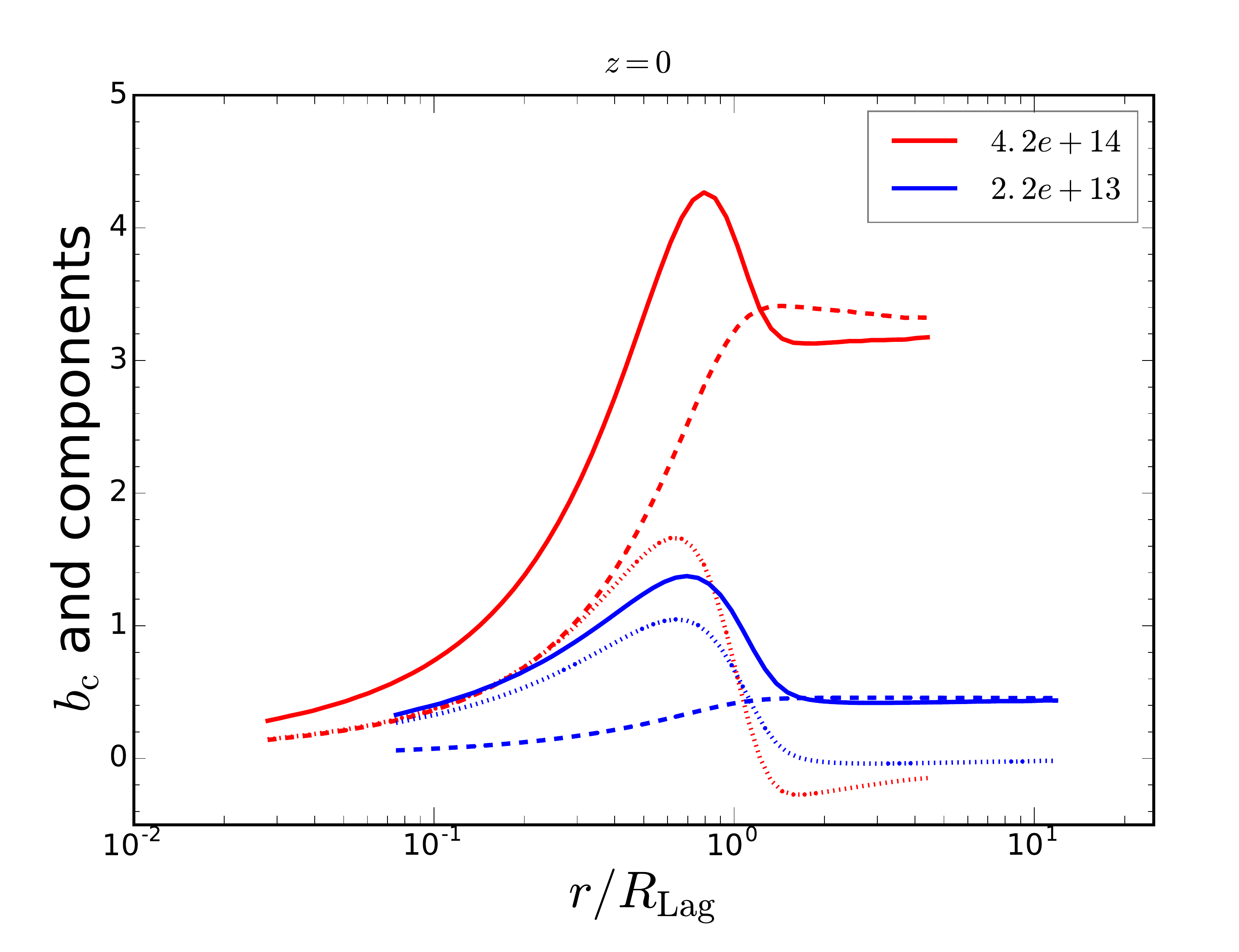}
\caption{ Configuration space cross-correlation signal, $b_c$, using best-fit parameters from the Fourier space fits shown in the main text, for two $z=0$ protohalo mass bins (as labelled).  Solid, dashed and dotted curves show the total signal, and the contributions which are proportional to $b_{10}$- and $b_{01}$-contributions. }
\label{fig:bc_component_model1}
\end{figure}

Fig.~\ref{fig:bc_TH_G_Eff} compares the cross bias parameter
\beq
 b_{\rm c}(r) \equiv \xi_{\rm c}(r)/\xi_{\rm m}(r)  
\eeq
for $W_{\rm TH}$, $W_{\rm G}$ and $W_{\rm Eff}$, when $R = 4 \MpcOh$, $b_{10} = 0.3$,  and $b_{01} = 0.8 $, which correspond to halos of mass $ 2 \times 10^{13} \Msun $ at $z=0$.  In all cases, the bias approaches $b_{10}$ on large scales ($r \gg R_{\rm Lag}$).  However, the contribution to $b_{\rm c} $ from the $b_{10}$-part drops rapidly as $\rho\lesssim  1$, while that from $b_{01}$ peaks sharply around $\rho \sim  1$. Therefore the bump in $b_{\rm c} $ is mainly driven by the term which is proportional to $b_{01}$. The sharpness of this bump depends on the smoothing window: it is least sharp for the Gaussian filter because $W_{\rm G}$ is substantially less compact than $W_{\rm TH}$. 

The other quantity which played an important role in the main text is the slope of the profile, 
$2\, d\xi_{\rm c}^W(r) / d \ln s_0(r)$, where $\xi_{\rm c}^W(r) $ is given by Eq.~\ref{eq:xic_WR}.  
For the tophat window
\begin{align}
\label{eq:xi2h}
  \frac{  d \xi_{\rm c}^{\rm TH }(r) }{ d \ln s_0(r)  } = 
  \begin{cases} \frac{3}{5} (b_{10}+ 6 b_{01})\, \xi_{\rm m}(R)\,\rho^2   &  \mbox{if } \rho\le 1    \\
     \frac{18 }{5}\,(b_{10}+b_{01})\, \frac{\xi_{\rm m}(R)}{\rho^3} 
    + \frac{3}{2}\, b_{10}\, \xi_{\rm m}(r)\, \left[1 - \frac{3}{\rho^2}\right] & \mbox{if } \rho > 1
              \end{cases} , 
\end{align}
where $\rho = r/R$, while for the Gaussian window 
\beq
 \frac{ d \xi_{\rm c}^{\rm G}( r ) }{ d \ln s_0(r)  }
 = \sqrt{ \frac{2}{\pi} } \xi_{\rm m}(r) \frac{\rho_{\rm G}^3 }{ (1 + \rho_{\rm G}^2)^{3/2} } 
     \Big[ b_{10} + \frac{6\,b_{01}}{1 + \rho_{\rm G}^2} \Big] , 
\eeq
with $\rho_{\rm G} = r / R_{\rm G}   $.   Fig.~\ref{fig:dxic_dlns0_2_TH_G_Eff} shows $2\, d\xi_{\rm c}^W(r) / d \ln s_0(r)$ for the three smoothing windows, using the same parameters as for Fig.~\ref{fig:bc_TH_G_Eff}.  All three curves show a peak near $ r \sim R$ ($R=R_{\rm G}$ for Gaussian). However, $W_{\rm TH}$ is narrower and cuspier than $W_{\rm G}$.  The curve for $W_{\rm Eff}$, with a narrow but rounded peak, is otherwise rather similar to that for $W_{\rm TH}$.  More importantly, it is qualitatively similar to those shown in Fig.~\ref{fig:u_rnorm} of the main text.

\begin{figure*}
\centering
\includegraphics[width=0.9\linewidth]{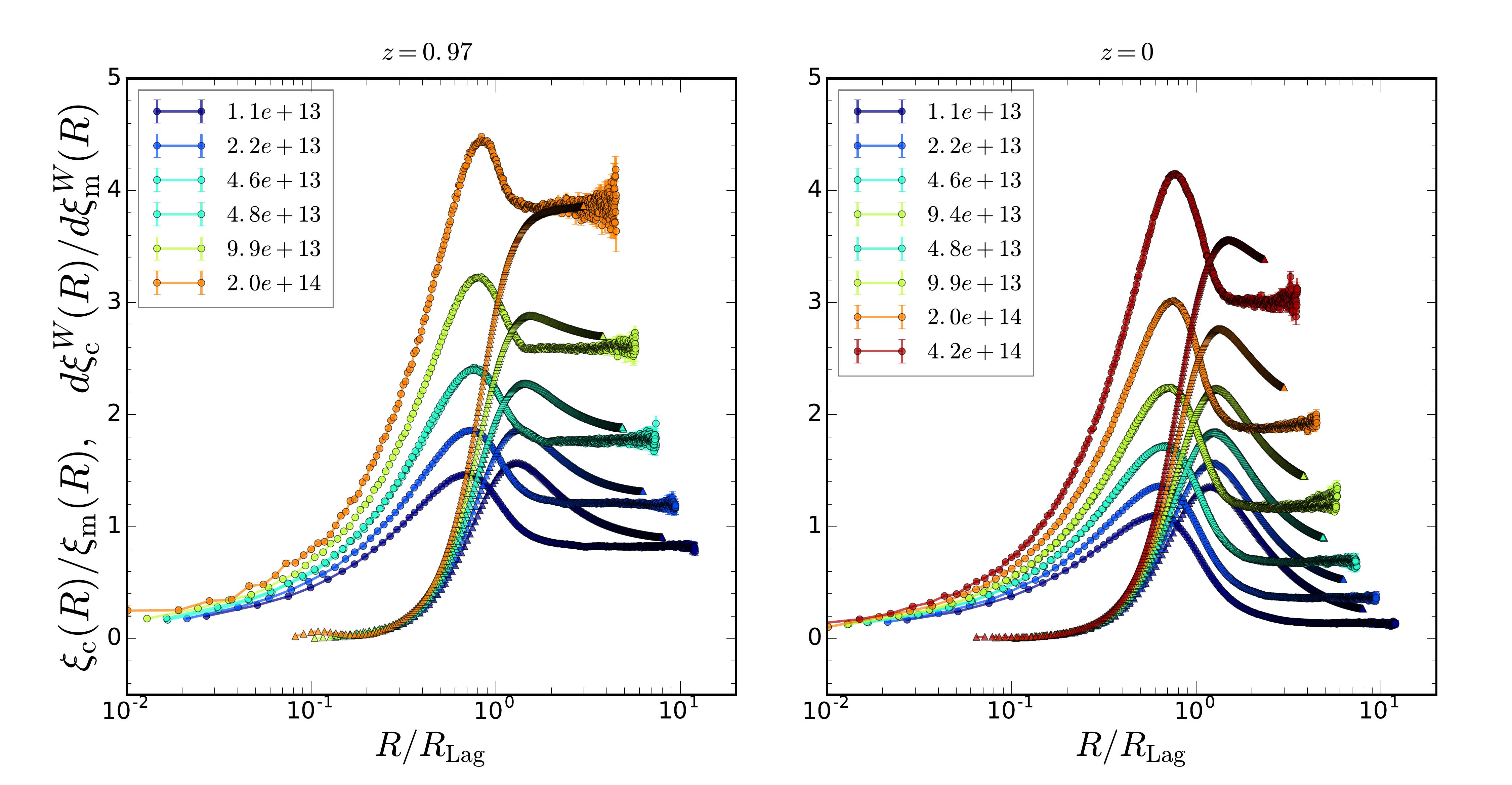}
\caption{Comparison of the ratio $\xi_{\rm c}/\xi_{\rm m}$ (circles) which is usually used to define the linear bias factor $b_c$ in configuration space, with $d\xi_{\rm c}^W/d\xi_{\rm m}^W$ (triangles)  as a function of $R$.     }
\label{fig:xic_xim_dxic_dxim}
\end{figure*}

\subsection{Lagrangian cross bias in configuration space}

The main text made extensive use of the bias parameters estimated from fitting to Fourier space measurements.  In principle, we could have fitted to configuration-space measurements instead.  Rather than doing so here, we instead show that the Fourier-space analysis is able to provide a good description of our configuration-space measurements.  

Fig.~\ref{fig:xic_rnorm} shows the configuration space cross-correlations for the same halo populations shown in Fig.~\ref{fig:bc_Lag_Dextrap}. From these, we defined the configuration space cross-bias parameter 
\beq
 \label{eq:bcL_r}
 b_{\rm c}( r, z) = \frac{ D(z_*) }{D(0)}\,
                    \bigg( \frac{ \xi_{\rm c}(r, z_*;z)  }{\xi_{\rm m}( r, z_*) } - 1  \bigg), 
\eeq
where $\xi_{\rm m}$ is the linear theory correlation function of dark matter at $z_*$ and $z$ is the redshift at which the Eulerian halos which we used to define the Lagrangian protohalos, were identified.  Comparison with Eq.~\ref{eq:bcL_k} shows that $b_{\rm c} (r,z)$ is defined similarly to $b_{\rm c} (k,z)$.  
Note that $\xi_{\rm m}(r, z_*)$ is computed, not from the particle distribution, but by performing the integral in Eq.~\ref{pk2xir} over the linear theory power spectrum $P_{\rm m}(k,z_*)$.

The symbols in Fig.~\ref{fig:bc_rnorm} show our measurements of $b_c$.  They are well described by the smooth curves which are not fits; rather, they show the predicted shape which we obtained by setting 
\beq
 \label{eq:bpk2bxir}
 \xi_{\rm c}^{\rm thy}(r) 
  = \int \frac{ dk\, k^2}{2\pi^2} \, b_{\rm eff}(k)\, P_{\rm m}(k,z=0)\,j_0(kr) , 
\eeq
where $b_{\rm eff }$ is given by Eq.~\ref{eq:bceff}, with $b_{10}$, $b_{01}$ and $R$ taken from Fig.~\ref{fig:bestfit_param_nu} (i.e., from fitting to the cross-spectrum bias shown in Fig.~\ref{fig:bc_Lag_Dextrap}), and then inserting this, instead of the measured $\xi_{\rm c}$ in Eq.~\ref{eq:bcL_r}.  Except for the most highly biased objects, these curves agree with the measurements reasonably well.  This demonstrates that our analysis does not suffer from large systematic effects associated with transforming between configuration and Fourier space.  

To examine the predictions in more detail, Fig.~\ref{fig:bc_component_model1} shows the $b_{10}$- and $b_{01}$-contributions (dashed and dotted) for two mass bins (as labelled) at $z =0$. For  $r /R_{\rm Lag} \gtrsim  1.5 $, the apparently constant part in fact has contributions of opposite signs, which cancel to yield the large-scale constant value. 
(Fig.~\ref{fig:bc_rnorm} suggests that high mass halos may suffer larger systematics at large $r$ than the low mass ones as deviations from the measurements are more apparent. However, it is the fractional deviation which matters, and these are comparable to or smaller than for the low mass halos.)  For  $r/R_{\rm Lag}\sim 1$, both components are positive.  As for the simple example shown in Fig.~\ref{fig:bc_TH_G_Eff}, the bump is primarily driven by the sharp rise of the $b_{01} $-component at  $r/R_{\rm Lag}  \sim  0.7 $.   

\subsection{Bias from correlating with large scale slope rather than density}
The main text defines $b_{\rm eff}(k)$ of Eq.~\ref{eq:bceff} as the ratio of the Fourier transform of 
$\langle \delta|{\rm protohalo}\rangle$ to $P_{\rm m}(k)$, where $\delta$ is the unsmoothed dark matter fluctuation at distance $r$ from the protohalo.  If $\Delta$ denotes $\delta$ smoothed with a filter $W(kR)$, then the Fourier transform of $\langle \Delta\delta\rangle$ equals $P_{\rm m}(k)\,W(kR)$, and the Fourier transform of $\langle \Delta|{\rm protohalo}\rangle$ divided by $P_{\rm m}(k)\,W(kR)$ also equals $b_{\rm eff}(k)$ of Eq.~\ref{eq:bceff}.  

With this in mind, consider Eq.~\ref{eq:xic_WR} for $\xi_c^W(R)$.  Writing $\xi_c(r)$ as the Fourier transform of $P_c(k) = b_{\rm eff}(k)\,P_{\rm m}(k)$, and rearranging the order of the integrals over $r$ and $k$ yields 
\beq
 \xi_{\rm c}^W(R) = 
 \int \frac{dk}{k}\frac{k^3 P_{\rm m}(k)}{2\pi^2}\,b_{\rm eff}(k)\, W(kR).
\eeq
Therefore, the slope variable 
\beq
 2\,\frac{d\xi_c^W(R)}{d\ln s_0(R)} = 
 \int \frac{dk}{k}\frac{k^3 P_{\rm m} (k)}{2\pi^2}\,b_{\rm eff}(k)\, 2\,\frac{dW(kR)}{d\ln s_0(R)}.
\eeq
Hence, if we define $\xi_{\rm m}^W$ and $d\xi_{\rm m}^W/d\ln s_0$ as the values of the expressions above when $b_{\rm eff}=1$, then we expect the ratio
\beq
 \frac{d\xi_{\rm c}^W(R)}{d\xi_{\rm m}^W(R)} = \frac{2\,d \xi_{\rm c}^W(R)/d\ln s_0(R)}{2\,d \xi_{\rm m}^W(R)/d\ln s_0(R)}
\eeq
to give $b_{10}$ on large scales. This is a special case of a more general point made by \cite{CastorinaParanjapeSheth2016} that, for any $Y_k$ which is linearly proportional to $\delta_k$, the Fourier space ratio $\langle Y|{\rm protohalo}\rangle/\langle Y_k\delta_k\rangle = b_{\rm eff}(k)$.  

Fig.~\ref{fig:xic_xim_dxic_dxim} compares  $d\xi_{\rm c}^W/d \xi_{\rm m}^W$ with $ \xi_{\rm c}/  \xi_{\rm m}$ using the same measurements as in Fig.~\ref{fig:bc_rnorm}. We find that $d \xi_{\rm c}^W / d \xi_{\rm m}^W$ indeed approaches $b_{10}$ on large scales, so that $d \xi_{\rm c}^W / d \xi_{\rm m}^W \approx \xi_{\rm c}/\xi_{\rm m}$ on large scales, in agreement with the assertion of \cite{CastorinaParanjapeSheth2016}.  

\subsection{Dependence on smoothing window:  Practice}
We noted in the main text that, although the correlations between variables which our methodology exploits may depend on a smoothing window $W$, it does not assume any particular functional form for $W$.  Our use of $W_{\rm Eff}$ in the main text was motivated by the fact that, with it, one obtains a good description of the cross-correlations between protohalos and the matter.  To illustrate that our methodology is not tied to this functional form, we now show the result of using $W_{\rm TH}$ instead.  Provided that using $W_{\rm TH}$ in Eq.~\ref{eq:bceff} does provide a good description of the curves shown in Fig.~\ref{fig:bc_Lag_Dextrap}, our methodology asserts that $b_{10}$, $b_{01}$, $s_0$ and $\xi_{\rm c}^{\rm W}$ may all change, but the consistency relation between them, Eq.~\ref{eq:ConsistencyRelation_2}, will still apply.

In practice, because $W_{\rm TH}\approx W_{\rm Eff}$ only at $k\ll 0.2 \, \hOMpc$ or so, we already know that a model based on $W_{\rm TH}$ will not be optimal, especially for determining $b_{01}$.  However, if we determine $b_{10}$ and $b_{01}$ from fitting to $k\le 0.2 \hOMpc$ only, then it is possible that everything will work as well as it did for $W_{\rm Eff}$.  In particular, since both $W_{\rm TH}$ and $W_{\rm Eff}\to 1$ on large scales, we expect $b_{10}^{\rm TH}\approx b_{10}^{\rm Eff}$.  If we set both $R_{\rm TH}$ and $R_{\rm Eff}$ equal to $R_{\rm Lag}$, we expect $b_{01}^{\rm TH}\approx (6/5)\,b_{10}^{\rm Eff}$ at low $k$. We can do it more accurately by matching  $b_{\rm eff}$ obtained using $W_{\rm TH}$ with that from  $W_{\rm Eff}$ up to second order.  These changes, along with the fact that $s_0^{\rm TH}(R_{\rm Lag})\ne s_0^{\rm Eff}(R_{\rm Lag})$, obviously impact one side of the consistency relation.  On the other hand, $\xi^{\rm TH}(R_{\rm Lag})\ne \xi^{\rm Eff}(R_{\rm Lag})$ either. In fact from Fig.~\ref{fig:dh_EffTH}, the threshold measured using the effective window is systematically lower than that from the top-hat. 
So it is possible that this change to the rhs of the consistency relation compensates to a large extent.  

\begin{figure}
\centering
\includegraphics[width=\linewidth]{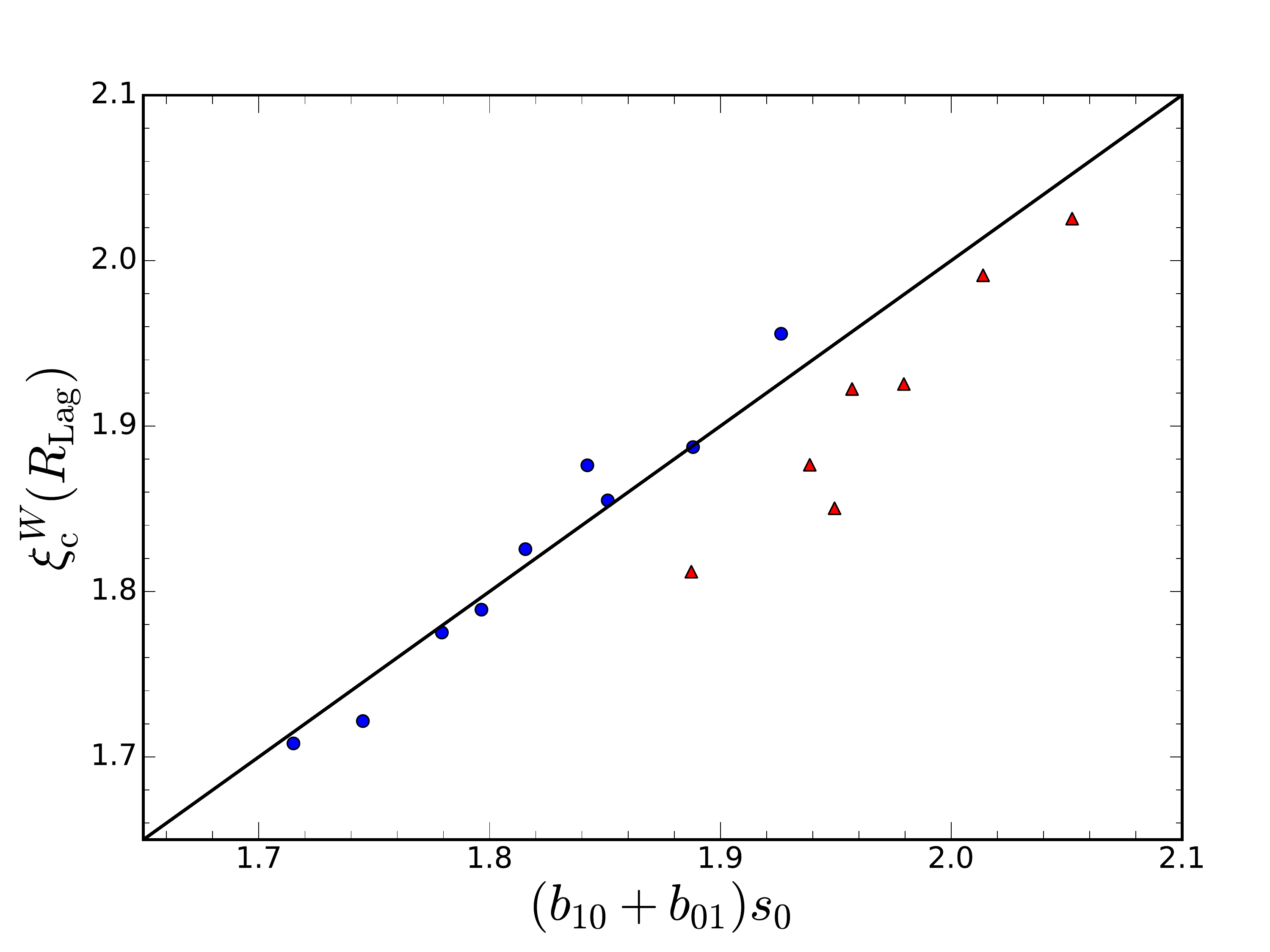}
\caption{ Dependence of the consistency relation for the enclosed density on the shape of the smoothing window.  Blue circles and red triangles use $W_{\rm Eff}$ and $W_{\rm TH}$ of Eqs.~\ref{eq:Weff} and~\ref{eq:Wth} when estimating $b_{10}$, $b_{01}$, $s_0$ and $\xi_{\rm c}^W$ for the protohalos of the $z=0.97$ and $z=0$ halos shown in Fig.~\ref{fig:bij_consistency_WeffPara_nu}.
}
\label{fig:compareWs}
\end{figure}

Fig.~\ref{fig:compareWs} shows that this is indeed what happens.  Blue circles and red triangles compare the two sides of the first equation in Eq.~\ref{eq:ConsistencyRelation_2} for $W_{\rm Eff}$ and $W_{\rm TH}$.  Note that this way of presenting the consistency relation test is different from that in Fig.~\ref{fig:bij_consistency_WeffPara_nu}, since there we also wanted to show the mass dependence of the enclosed overdensity, whereas here the object of interest is the consistency relation itself.  Both sets of symbols in Fig.~\ref{fig:compareWs} lie close to the one-to-one line, indicating that the consistency relation applies both for $W_{\rm Eff}$ and for $W_{\rm TH}$.  

\begin{figure}
\centering
 \includegraphics[width=\linewidth]{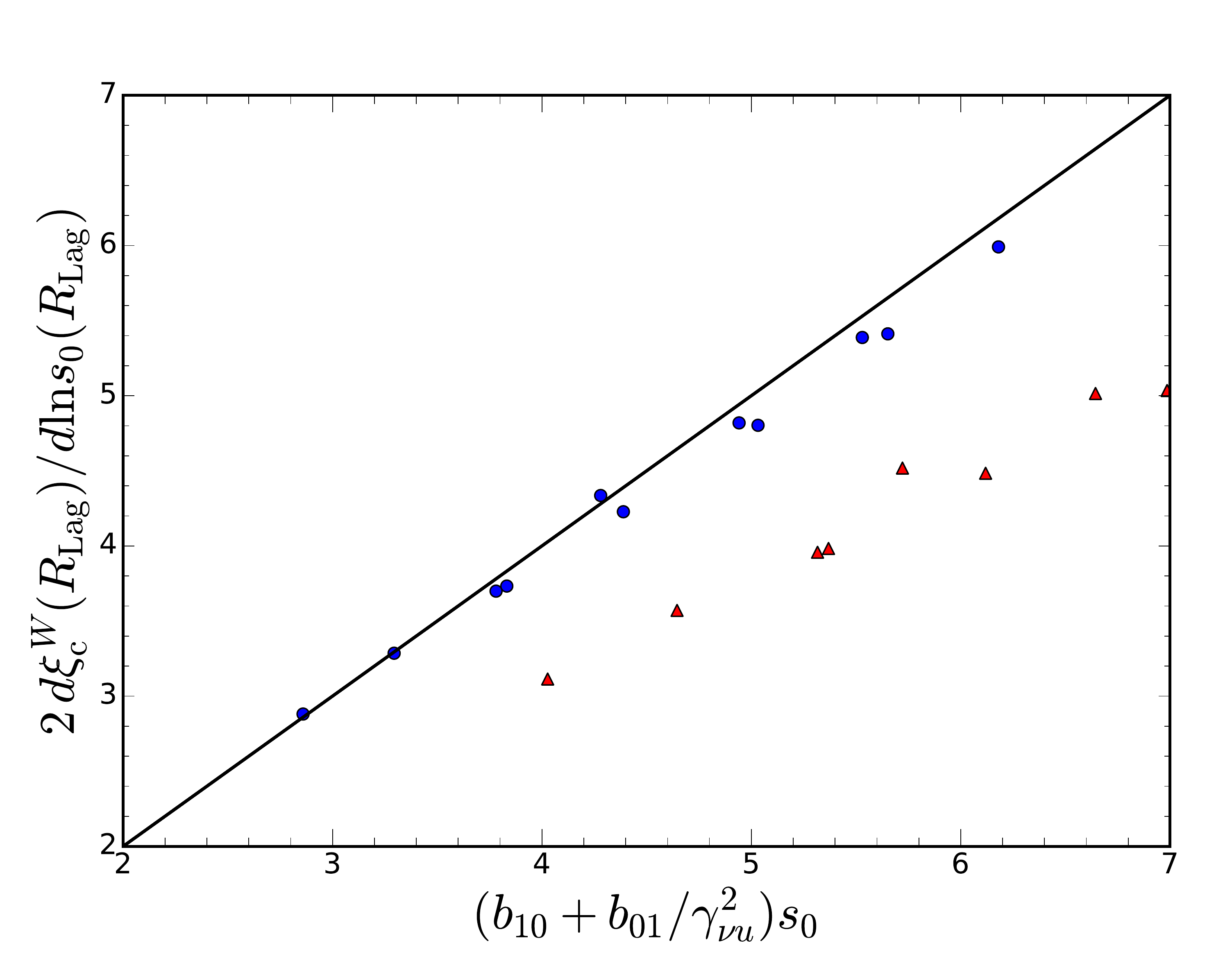}
 \caption{Same as Fig.~\ref{fig:compareWs}, but now for the consistency relation associated with the profile slope.}
\label{fig:slopeWs}
\end{figure}

\begin{figure}
\centering
\includegraphics[width=\linewidth]{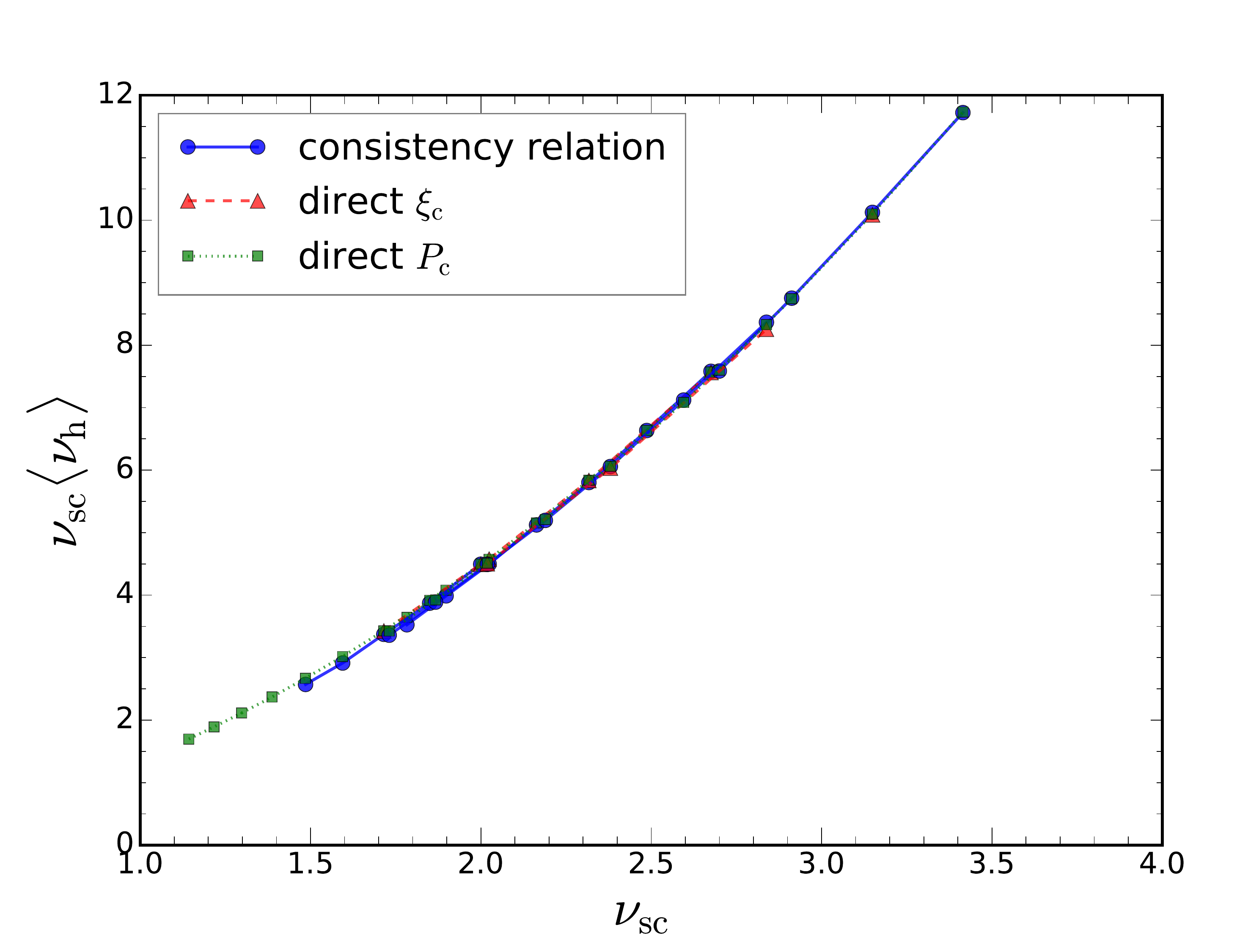}
\caption{Test of the first consistency relation (in Eq.~\ref{eq:ConsistencyRelation_2}).  Blue circles connected with solid curves show $(b_{10} + b_{01}) \delta_{\rm sc}(z)$.  Red triangles (dashed)  and green squares (dotted)  show $\langle\delta_{\rm h}\rangle\, \delta_{\rm sc}(z)/s_0$, where $\langle\delta_{\rm h}\rangle$  comes from direct measurements of the enclosed overdensity in protohalo patches using real and Fourier-space methods (Eq.~\ref{eq:xic_WR} and~\ref{eq:Pk2dh}) respectively. The data  from the halos of mass $M$  at $z=0$  and 0.97 in the Oriana and Carmen simulations, are shown as a function of the scaled mass variable $\nu_{\rm sc}=\delta_{\rm sc}(z)/\sqrt{s_0(M)}$. }
\label{fig:cheat}
\end{figure}

The agreement is slightly worse for $W_{\rm TH}$, presumably because using $W_{\rm Eff}$ in Eq.~\ref{eq:bceff} is simply a better model for the cross bias \citep{ChanShethScoccimarro2015}.  (For $W_{\rm TH}$, we have actually refit $b_{10}$, $b_{01}$ and $R$, rather than rescaling as described above.  The resulting fractional deviation in Fig.~\ref{fig:compareWs} is of the same order as $s_0^{\rm TH}$ from $s_0^{\rm Eff}$.)  Fig.~\ref{fig:slopeWs} shows a similar analysis of the consistency relation for the slope.  In this case, $W_{\rm Eff}$ fares substantially better; presumably this is because, as Fig.~\ref{fig:dxic_dlns0_2_TH_G_Eff} shows, $W_{\rm TH}$ results in a cuspy signal around $R_{\rm Lag}$, so it changes significantly if the appropriate scale differs even slightly from $R_{\rm Lag}$, which can happen if $b_{\rm eff}$ is not a perfect fit to $b_c$.  

Recently, \cite{Modi:2016dah} report the results of a similar test of the consistency relation for $\langle \delta_{\rm h} \rangle$ based on $W_{\rm TH}$.  However, they presented their results in a format which makes it difficult to assess ten percent discrepancies.  Our Fig.~\ref{fig:cheat} shows our results in the format they used.  The agreement between the direct and large-scale structure measurements is impressive; however, the larger dynamic range here does not do justice to the level of agreement which our preferred formats, Figs.~\ref{fig:bij_consistency_WeffPara_nu} and~\ref{fig:compareWs}, show.  Unfortunately, they do not show results for the slope.  Nevertheless, the agreement between our analyses for $\langle\delta_{\rm h}\rangle$ is reassuring, as it indicates that detailed a priori knowledge of the shape of the smoothing window is not crucial for reconstructing $\langle\delta_{\rm h} \rangle$ from large scale bias -- in agreement with the arguments given in the main text.  

\bsp	
\label{lastpage}
\end{document}